\documentclass[journal,twoside,web]{IEEEtran}
\usepackage{tmi}
\usepackage{cite}
\usepackage{amsmath,amssymb,amsfonts}
\usepackage{algorithmic}
\usepackage{graphicx}
\usepackage{textcomp}
\usepackage{placeins}
\usepackage{hyperref}

\usepackage{listings}
\usepackage{gensymb}
\usepackage[ruled,vlined]{algorithm2e} 

\def\BibTeX{{\rm B\kern-.05em{\sc i\kern-.025em b}\kern-.08em
    T\kern-.1667em\lower.7ex\hbox{E}\kern-.125emX}}
\begin{document}
\title{Unsupervised Domain Adaptation from Axial to \\ Short-Axis Multi-Slice Cardiac MR Images by \\ Incorporating Pretrained Task Networks}

\author{Sven Koehler, Tarique Hussain, Zach Blair, Tyler Huffaker, Florian Ritzmann, Animesh Tandon, Thomas~Pickardt, Samir~Sarikouch, Heiner Latus, Gerald Greil, Ivo Wolf, Sandy~Engelhardt

\thanks{© 2021 IEEE. Manuscript accepted at TMI (Transaction on Medical Imaging): January 13, 2021.
This work was supported in part by Informatics for Life funded by the Klaus Tschira Foundation and by the Competence Network for Congenital Heart Defects, which has received funding from the Federal Ministry of Education and Research, grant number 01GI0601 (until 2014), and the DZHK (German Centre for Cardiovascular Research; as of 2015).}
\thanks{S. Koehler, F. Ritzmann, S. Engelhardt are with the Department of Internal Medicine III, Group Artificial Intelligence in Cardiovascular Medicine, Heidelberg University Hospital, D-69120 Heidelberg, Germany; DZHK (German Centre for Cardiovascular Research), Heidelberg/Mannheim, Germany.}
\thanks{T. Hussain, A. Tandon, G. Greil are with the Department of Pediatrics, Division of Cardiology; Department of Radiology; Advanced Imaging Research Center, UT Southwestern Medical Center, Dallas, TX, USA.}
\thanks{Z. Blair, T. Huffaker are with the Department of Pediatrics, UT Southwestern Medical Center, Dallas, TX 75390, USA.}
\thanks{T. Pickardt, S. Sarikouch are with the German Competence Network for Congenital Heart Defects; DZHK (German Centre for
Cardiovascular Research), Berlin, Germany.}
\thanks{S. Sarikouch is with the Department of Cardiothoracic, Transplantation and Vascular Surgery,
Hannover Medical School, Germany.}
\thanks{H. Latus is with the Department of Paediatric Cardiology and Congenital Heart Defects,
German Heart Centre Munich, Germany.}
\thanks{I. Wolf is with the Department of Computer Science, Mannheim University of Applied Science, Germany.}}

\maketitle

\begin{abstract}
Anisotropic multi-slice Cardiac Magnetic Resonance (CMR) Images are conventionally acquired in patient-specific short-axis (SAX) orientation. In specific cardiovascular diseases that affect right ventricular (RV) morphology, acquisitions in standard axial (AX) orientation are preferred by some investigators, due to potential superiority in RV volume measurement for treatment planning. Unfortunately, due to the rare occurrence of these diseases, data in this domain is scarce. Recent research in deep learning-based methods mainly focused on SAX CMR images and they had proven to be very successful. In this work, we show that there is a considerable domain shift between AX and SAX images, and therefore, direct application of existing models yield sub-optimal results on AX samples. We propose a novel unsupervised domain adaptation approach, which uses task-related probabilities in an attention mechanism. Beyond that, cycle consistency is imposed on the learned patient-individual 3D rigid transformation to improve stability when automatically re-sampling the AX images to SAX orientations. The network was trained on 122 registered 3D AX-SAX CMR volume pairs from a  multi-centric patient cohort. A mean 3D Dice of $0.86\pm{0.06}$ for the left ventricle, $0.65\pm{0.08}$ for the myocardium, and $0.77\pm{0.10}$ for the right ventricle could be achieved. This is an improvement of $25\%$ in Dice for RV in comparison to direct application on axial slices. 
To conclude, our pre-trained task module has neither seen CMR images nor labels from the target domain, but is able to segment them after the domain gap is reduced.
Code: \url{https://github.com/Cardio-AI/3d-mri-domain-adaptation}
\end{abstract}


\section{Introduction}
\label{sec:introduction}

\IEEEPARstart{A} major challenge in medical imaging and deep learning is, besides the rather small size of data sets and the scarce availability of corresponding labels,
constituted by the so-called data distribution shift.
In fact, different images (e.g., varying in slice positioning and orientation, spacing, machine vendor) can result in vastly different data distributions, despite imaging the same subject.
As a result, the performance of deep learning methods often degrades by a large margin if data distributions during training and testing do not match, meaning that the model was not properly adapted to the desired test  conditions. 

\textit{Domain Adaptation} (DA) is commonly referred to as an approach to address such a domain shift between a \textit{source} and \textit{target} distribution \cite{PanSurvey}. Considering a source domain with input images $X_{s}$ and distribution $P(X_{s})$ and $P(Y|X_{s})$, $Y$ being the labels, as well as a target domain with input image distribution $P(X_{t})$ and $P(Y|X_{t})$, $P(X_{s})\neq P(X_{t})$. Then, DA can be addressed via a supervised approach where annotated data from the target domain is available and incorporated into the approach, or via unsupervised learning where the target domain is unlabeled, meaning that $P(Y|X_{t})$ is not available \cite{toldo2020unsupervised,wang2018deep}.
The latter case reflects the most useful real-world scenario and is commonly referred to as unsupervised DA. 

The involved task is to leverage knowledge from the target domain using the unlabeled data available in $P(X_{t})$.
In this work, we propose an unsupervised DA approach that can be employed for 2D multi-slice stack Cardiac Magnetic Resonance (CMR) acquisitions with varying orientations. 

CMR imaging is a widely used modality to capture the anatomy of the heart and great vessels, to quantify ventricular size and function and to characterize myocardial tissue. 
For morphological and volumetric assessment, specific CMR sequences such as cine-SSFP (steady-state free precession) images are preferably employed. 
Cine-SSFPs are bright-blood images, characterized by a high contrast between the myocardium and blood pool. 

In clinical routine, it is a well-established standard to capture short axis (SAX) views, which are perpendicular to the septum of the heart and composed of several slices, also referred to as  multi-slice stack. The purpose of acquiring not thorax-aligned image orientations, is to have a direct reference between the image slices and the cardiac anatomy, as the orientation of the heart within the thorax varies between individuals. In consequence, the orientation of the SAX image stack must be manually defined during acquisition.

SAX images are useful in recovering the three-dimensional shape of the ventricles, e.g. for volumetric assessment and quantification of global pump function after contouring the endocardial border.
However, acquired slices are usually thick (typically about 7-9mm) compared to the in-plane resolution (typically 1-2mm) due to limited signal-to-noise ratio (SNR). 
On the other hand, nearly isotropic resolution is infeasible in clinical routine to be performed in patients with cardiovascular diseases due to long breath-hold and repetition times.

\begin{figure}
  \includegraphics[width=\linewidth]{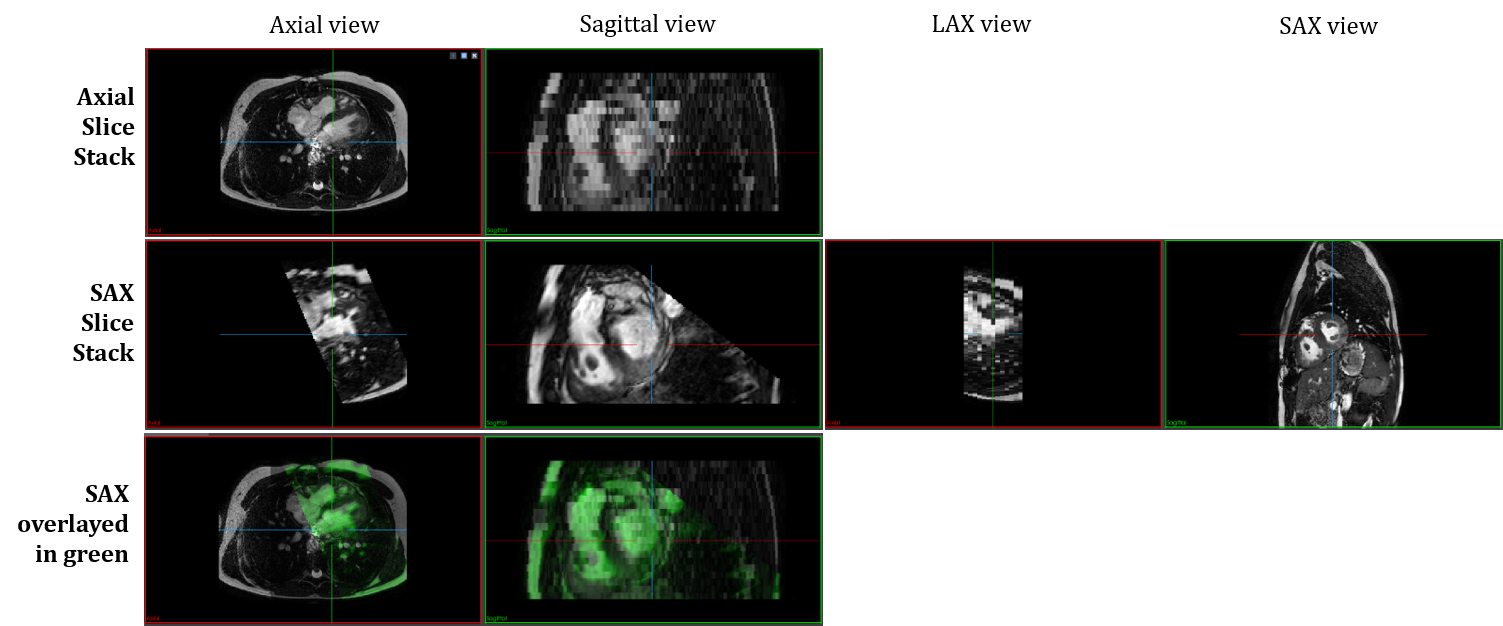}
  \caption{Comparison of axial (AX) and short axis (SAX) orientations of two SSFP slice stacks from the same patient in the same physical space. Note that for visualization purposes, SAX image is resampled in cubic interpolation mode in axial and sagittal view. In the other cases, the differences in voxel spacings are clearly visible.}
  \label{fig:comp}
\end{figure}

\begin{figure}
  \includegraphics[width=\linewidth]{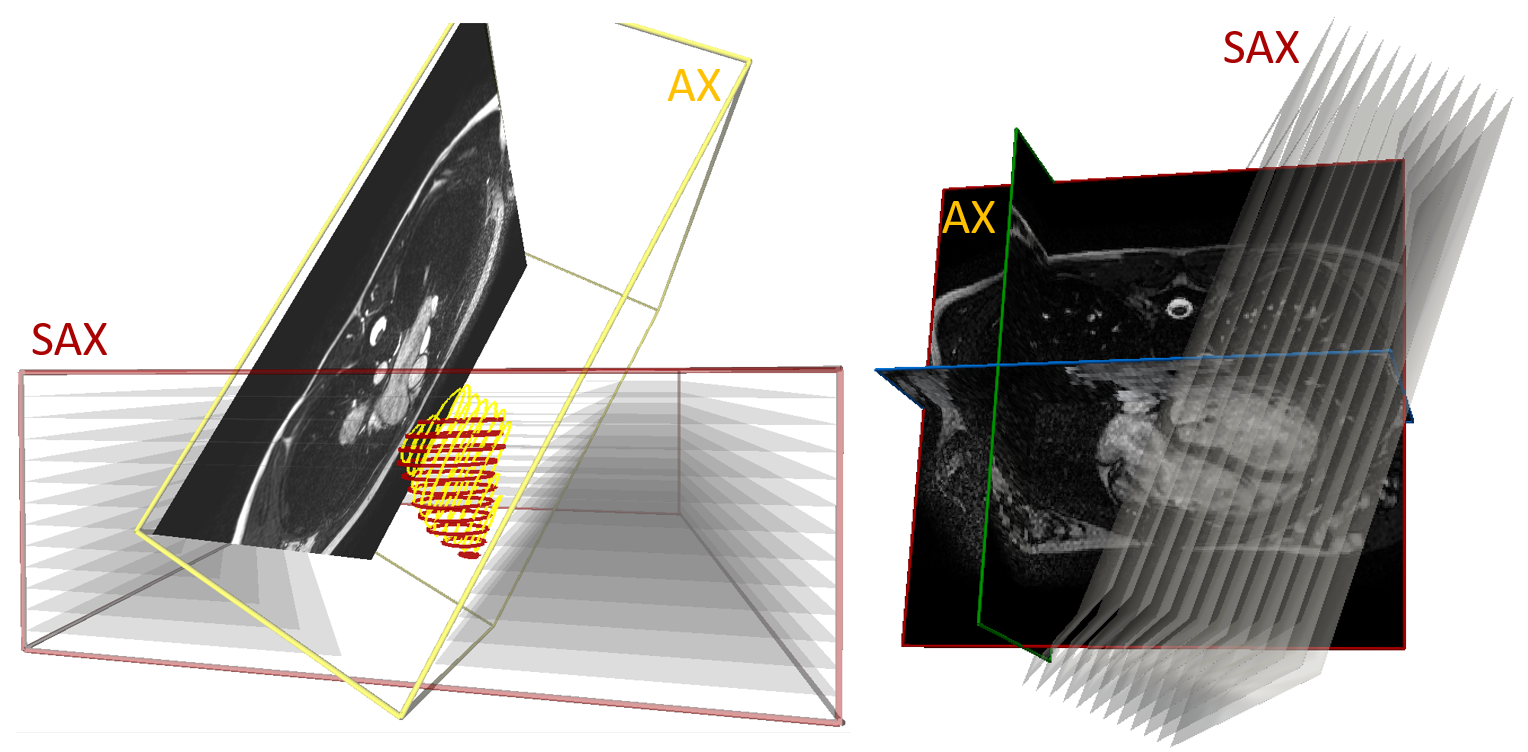}
  \caption{3D-Spatial relation between SAX (red) and AX (yellow) images. Example contours from left ventricle are shown.}
  \label{fig:spatialrelation}
\end{figure}

In specific cardiovascular diseases that affect right ventricular (RV) geometry, however, thorax-aligned axial (AX) slices (Fig.\ref{fig:comp} and Fig.\ref{fig:spatialrelation}) are preferred  by some investigators as some studies reported superior reproducibility of RV volume measurements compared to SAX slices \cite{Fratz2009ComparisonOA}. This is especially important for patients with abnormal RV morphology such as arrhythmogenic right ventricular cardiomyopathy (ARVC) and patients with congenital heart defects (CHD) such as repaired tetralogy of fallot (TOF) \cite{Fratz2009ComparisonOA,Fratz2009} and Ebstein’s anomaly \cite{Fratz2009}. In patients with CHD, the RV is frequently subject to chronic volume or pressure overload and quantification of RV volumes is crucial for individual treatment planning, evaluation of response to therapy and estimation of prognosis.

Yet, there is hardly any work on deep learning approaches for AX CMR. 
This might be due to the fact that available data sets and corresponding annotations are too scarce for training competitive fully convolutional networks (FCN) from scratch.
Therefore, it would be highly beneficial to exploit existing labels from related acquisitions as efficiently as possible \cite{TajbakhshMIA}.

In this work, we show that existing models trained on SAX slices are not directly applicable to AX slices, even if they were trained on a database which includes the same pathology. To reduce the shift between these two domains, we propose an unsupervised domain adaptation approach that is able to learn the geometric patient-specific relation between AX and SAX by regressing on the position and orientation parameters of the slice stacks using an image-based loss. 

The actual goal is to re-sample an existing volume with a different orientation and translation into a target grid of fixed size to obtain a SAX view. The model is allowed to translate and rotate the original image with 6 degrees of freedom. In particular, the axial acquisitions are re-sampled to a SAX orientation by a rigid affine transformation predicted via a novel transformation module. Then, a task with pixel-wise predictions can be performed on the SAX representation and the resulting image is transformed back to the target domain. 

In summary, the contributions of this paper are as follows:
\begin{itemize}
    \item \textit{Unsupervised Domain Adaptation(C1):} We show how to efficiently reuse existing task networks trained on SAX CMR databases to solve the same task (here: segmentation of left ventricle, LV; right ventricle, RV; and myocardium, MYO) on the target domain, namely axial acquisitions. The approach does not require labels from the target domain and is therefore unsupervised and more generally applicable than simple data fusion techniques.
    \item \textit{Weak Dependency on Task Network (C2):}  Our DA approach incorporates other segmentation networks without requiring further adjustments, and is only weakly coupled to an associated task due to the task-independent loss.
    \item \textit{Extended Transformation Module (C3)}: To bridge the most obvious domain gap between AX and SAX samples, we enable the model to learn the 3D geometric relations between the two domains. The performance of the transformer is improved by the following two contributions:
    \begin{itemize}
    \item \textit{Cycle-Constraints:}
     Cycle-consistency in AX-to-SAX and SAX-to-AX transformations is enforced and exploiting this relation enables to transform \textit{from} target \textit{to} source domain \textit{and} back.
    \item \textit{Task-Specific Focus:} By incorporating an existing task-specific model trained on the source domain, we propose to use the inferred voxel-wise probabilities in the loss function to guide the transformation module for including relevant regions essential to reliably solve the task.  
    
    \end{itemize}
    \item We evaluate the performance of our method components on a series of experiments against a baseline, using a heterogeneous multi-center population of TOF patients.
\end{itemize}

\section{Related Work}

\subsection{Cine-SSFPs Segmentation}

Since the introduction of FCN-based models for segmentation, in particular the U-Net \cite{Ronneberger2015}, the community could witness remarkable progress in high-performing segmentation models trained on cine SSFPs images on SAX \cite{Isensee2018,Bernard2018a,ACNN} or on SAX and long-axis (LAX) combined \cite{OmegaNet2018}.
Especially the challenges held in the last years serve as a great benchmark of state-of-the-art methods; to list a few examples: 
the \textit{Left Ventricle Segmentation Challenge} (LVSC at MICCAI 2011) \cite{LVSC}, the \textit{Right Ventricle Segmentation Challenge} (RVSC at MICCAI 2012) \cite{RVSC_Miccai2012}, the \textit{Automated Cardiac Diagnosis Challenge} (ACDC at MICCAI 2017) \cite{Bernard2018a}. 

With the advent of the deep learning era, the 8 top performing algorithms in the ACDC segmentation sub-challenge were based on convolutional neural networks (CNNs). For example, the work by Isensee \textit{et al.}~\cite{Isensee2018} proposed a 2D-/3D-U-Net ensemble approach. According to \cite{Bernard2018a}, this approach is closely followed by other methods which are less than one pixel away from  it,  especially  for  the  left ventricle (LV)  and  RV  at  end-diastole (ED). One of those methods is the method by Baumgartner \textit{et al.}~\cite{BaumgartnerACDC}, who used a 2D U-Net with cross-entropy loss that performed best in their trials in comparison to a 3D U-Net. In fact, most of the submissions dealt with a 2D U-Net solution \cite{Bernard2018a}, due to the highly anisotropic nature of the data and the higher amount of training data in slice-based approaches. 
Compared to the traditional level-set methods (e.g. \cite{Grinias_ACDC17}), the submitted deep learning methods achieved considerably higher accuracy.
However, all of these challenges and the presented approaches solely addressed segmentation tasks on SAX multi-slice acquisitions. While the challenges are a valuable source for freely available data, data acquired in axial orientation is scarce and labels are not directly available.

\subsection{Spatial Transformer Networks}
\label{sec:spatialtransformer}
Jaderberg \textit{et al.}\cite{STN_Nips} introduced a novel differentiable module called  Spatial Transformer Networks (STN),   
which explicitly allows the spatial manipulation of images within a CNN.
The authors showed that it can be employed to learn model invariance to translation, rotation, scale and  non-rigid deformations. Relevant image regions can be identified, and transformed to a canonical, expected pose to simplify the task. 
$\Omega$-Net \cite{OmegaNet2018} employs this concept to transform LAX and SAX slices into a canonical orientation in 2D. The proposed method generalizes to images with varying sizes and  orientations, however, alignment in 3D was not addressed. 
Other work proposed to use the flow field representation of STNs for deformable 3D image registration \cite{Balakrishnan2018}. The same group later \cite{Dalca2019UnsupervisedLO} integrated a diffeomorphic layer, enforcing an inverse-consistent constraint on the deformation fields. 
In \cite{Kim2019}, Kim \textit{et al.} extended on this by incorporating a cycle-consistency into the registration problem. Inspired by this, our work includes rigid affine transformations and their inverse to exploit cycle-consistency during the learning process.

\subsection{Unsupervised Domain Adaptation}

The recent review paper by Toldo \textit{et al.} \cite{toldo2020unsupervised} divided the research field of unsupervised DA (UDA) in semantic segmentation into seven ongoing directions. 
First, Weakly and Semi-Supervised learning is in particular relevant for domains with expensive, heterogeneous and sparse labelled data.
Recent work \cite{papandreou2015weakly,dai2015boxsup,kolesnikov2016seed} showed that segmentation models could benefit from weak labels. 
Second, the usage of adversarial training (e.g., \cite{goodfellow2014generative}) for UDA showed impressive results for input-, feature- and output adaptation \cite{chen2018road, chen2019crdoco, hoffman2018cycada}, which is related to our approach, where we try to adopt the input space.
Third, generative-based approaches with unpaired data samples were recently reported with CycleGANs \cite{li2019bidirectional, hoffman2018cycada, chen2019crdoco, toldo2020unsupervised}; these approaches align with ours in the idea of exploiting forward and backward transformation in a cycle consistency manner.
Forth, Saito \textit{et al.} \cite{saito2017adversarial, saito2018maximum} proposed a method to overcome the issue of ignored decision boundaries within the generative based approaches by incorporating Adversarial Dropout Regularization. This field is called Classifier Discrepancy.
Fifth, recently Self-Training has also been applied for UDA \cite{zou2019confidence}. In their work, they used the segmentation network output as a measure of reliability for generated labels which were then used as new ground truth to further exploit the unknown domain \cite{michieli2020adversarial, spadotto2020unsupervised}. This idea is related to our $Loss_{focus}$ extension. 
The last two remaining directions, Curriculum Learning and Multi Tasking, are both trying to incorporate additional easier-to-learn tasks, and, if available, privilege information to further overcome the distribution shift towards unknown domains \cite{dai2020curriculum, lee2018spigan}.

These directions can be further grouped by the possible adaptation levels (input, feature, output adaptation) \cite{toldo2020unsupervised}. The first two directions try to tackle the problem on the input level, directions four and six avoid dealing with an excessively convoluted latent space and tackle the cross-domain distribution over the segmentation output space. The other four directions seek for an alignment of the latent network embedding forcing the feature extractor to discover cross-domain generalised features. Our approach belongs to the first group of adaptation levels as we transform the target images to the source domain which are then fed to the segmentation module.

\section{Material and Methods}

\begin{figure}
\centering
 \includegraphics[width=\linewidth]{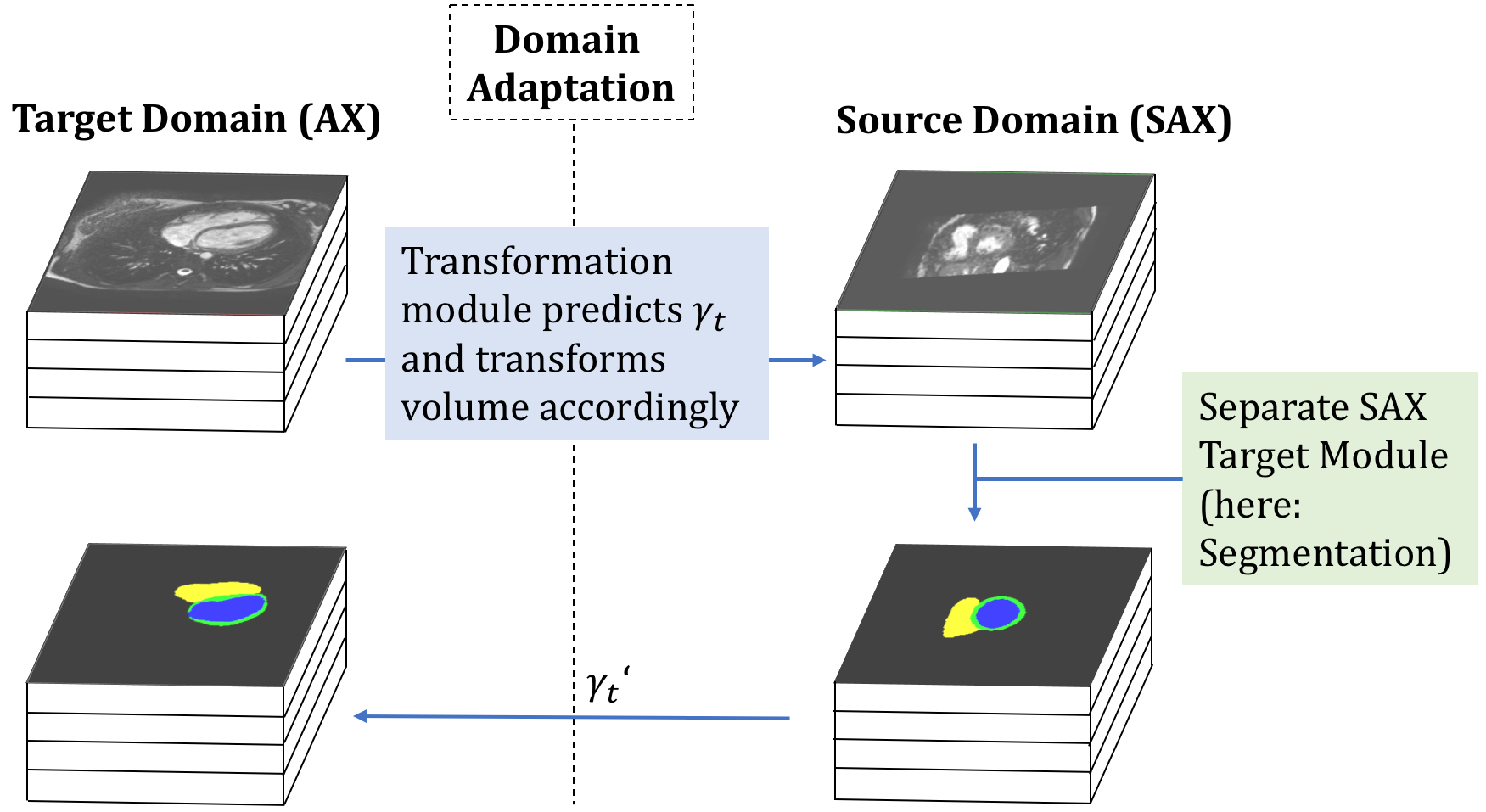}
  \caption{The application scenario, the target image is transformed according to the estimated transformation parameters $\gamma_{t}$; a task, like a segmentation is performed and the resulting mask is transformed back into the target domain.}
\label{fig:application}
\end{figure}

Our application scenario is presented in Fig. \ref{fig:application}. 
Given data in an orientation with only scarce representatives, like axially orientated CMR image stacks, a trainable transformation module transforms the data to the prevalent orientation, in our case the SAX domain. Now, an existing task network, trained on samples in the prevalent orientation, such as a U-Net segmentation module trained on SAX images, is applied and the result is subsequently transformed back into the original domain.

\subsection{Example Task Network: Segmentation Module}
\label{sec:unet}

\begin{figure}
\centering
  \includegraphics[width=\linewidth]{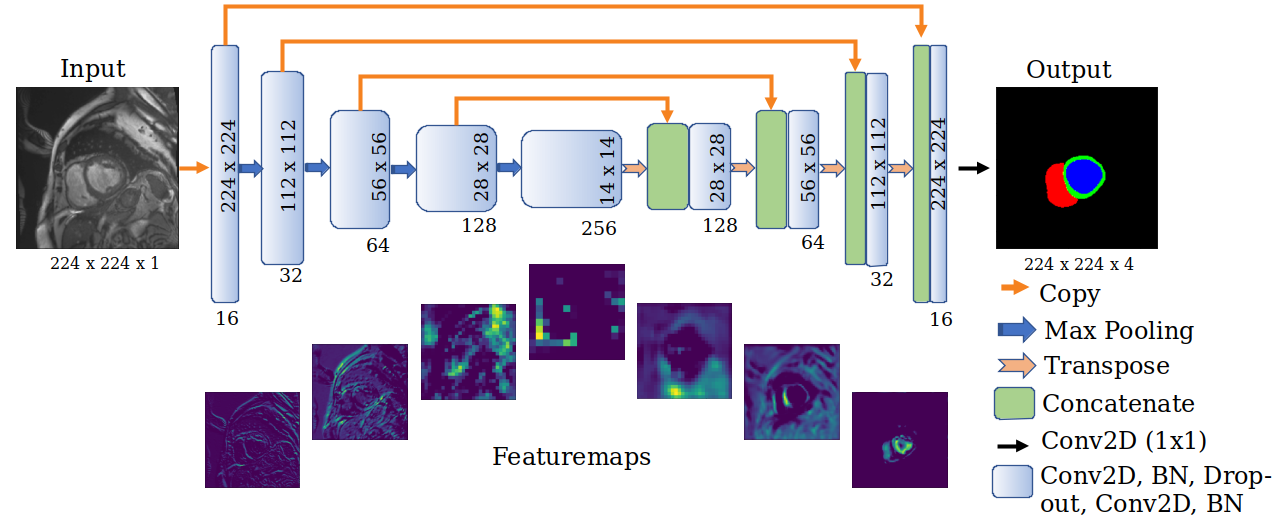}
  \caption{U-Net based segmentation network architecture.}
  \label{fig:unet_overview}
\end{figure}

The aim of the segmentation module is to delineate the voxels of three different classes, which are the left ventricle (LV), the right ventricle (RV) and the left ventricular myocardium (MYO)  from the background.
For solving this task, a 2D U-Net architecture (Fig. \ref{fig:unet_overview}) was chosen based on the original architecture of Ronneberger \textit{et al.} \cite{Ronneberger2015}. The hyper-parameters of this network were investigated and optimised with regard to their generalisation capabilities in a previous study \cite{Koehler2020}.

At that time minor adjustments were made, such as the use of Exponential Linear Unit (ELU) activation functions, defined by
\begin{align}
\label{eq:elu}
f(x) = \left \{ \begin{array}{rcl} 
x                   & \mbox{for} & x > 0\\  
\alpha (e^{x} - 1)  & \mbox{for} & x \le 0  \end{array} \right.
\end{align}
with $\alpha = 1.0$, instead of Rectified Linear Units (RELU) activation, a batch normalization layer \cite{ioffe2015batch} after each non-linear ELU activation, higher dropout rates and a combined loss function $Loss_{\text{seg}}$, consisting of a cross-entropy ($CE$) and a Soft Dice loss ($SDL$). The binary cross-entropy $BCE$ is defined for each foreground class $j$, $j \in \{\text{left ventricle, right ventricle, myocard}\}$ as
\begin{align}
    BCE(Q_j,G_j) &= -\frac{1}{N} \sum_{i=1}^{N} \big( g_{ij} \cdot log (q_{ij}) \nonumber \\ &
    +(1 - g_{ij})\cdot log(1 - q_{ij}) \big), \label{eq:bce}
\end{align}
where the sum is calculated over the $N$ voxels of the softmax
output $q_{ij} \in [0,1]$ 
of the $i$-th pixel 
for foreground class $j$ and the corresponding ground truth binary mask volume $G_j$ with $g_{ij} \in \{0,1\}$. The cross entropy loss then sums up to
\begin{equation}
CE(Q,G) = \frac{1}{C}\sum_{j=1}^{C} BCE (Q_j,G_j)
\end{equation}
with $C$ being the number of foreground classes. The $SDL$ incorporates the binary soft Dice coefficient $DSC_j$ for each foreground class and is defined as 
\begin{align}
\begin{split}
  DSC(Q_j,G_j)     & = \frac{2 \cdot \sum_{i=1}^{N} g_{ij} \cdot q_{ij}  + smooth}{\sum_{i=1}^{N} g_{ij} + \sum_{i=1}^{N}q_{ij} + smooth}\label{eq:dice_class}
\end{split}
\end{align}
with $smooth$ is commonly set to 1, which among others avoids division by zero. 

Following the definition by Drozdzal \textit{et al.}\ \cite{drozdzal_importance_2016}, the $SDL$ is defined as   
\begin{align}
    SDL(Q,G) = 1 - \frac{1}{C} \sum_{j=1}^{C} DSC(Q_j, G_j).\label{eq:dsc_loss}
\end{align}

Finally, the segmentation loss is composed of 
\begin{align}
Loss_{seg} =  w_{seg} \cdot  CE(Q,G) + SDL(Q,G),
\end{align}
with $w_{seg} = 0.5$, which achieved the best performance investigated while working on \cite{Koehler2020}. To ensure that this weighting is also the best choice for the dataset used in this work, we evaluated the influence of different $w_{seg}$ values on the results within the scope of a sensitivity analysis (cf. Table \ref{tab:unetlossalpha}). 

After the prediction, images are post-processed using three-dimensional largest connected component filtering and two-dimensional morphological closing with a kernel size of $5\times5$ for each label.

\subsection{Transformation module (baseline)}
\label{sec:trans_base}
Our goal is to learn a 3-dimensional rigid transformation $\mathcal{T}_{\gamma}$ between the axial image $I \in \mathbb{R}^{H \times W \times D}$, which serves as the source image, and the transformed axial image $\hat{I}=\mathcal{T}(I,\hat{M})$. The relative transformation $\hat{M}$ between axial and short-axis CMR image pair can be computed from the respective tags in the DICOM header. 
Note that AX and SAX volume stacks describe a different physical volume, which means that each of the images contains information which is not present in the other image, as exemplary shown in Fig. \ref{fig:spatialrelation}. Therefore, our goal was not to learn the short-axis image directly, but a transformed axial image.
$W, H, D$, correspond to width, height and depth of the images, respectively.
In terms of notation, a superposed circumflex (e.g., $\hat I$) indicates ground truth in the following. 

The original spatial transformer network proposed by Jaderberg \textit{et al.} \cite{STN_Nips} consists of three main sub-modules: a localisation network, which predicts the parameters of a  transformation matrix, $M$;  the grid generator, which implements the transform, $\mathcal{T}$; and the sampler, which is responsible for the interpolation. 
The localisation network learns the parameters $\gamma=f_{\textrm{loc}}(I)$ of the transformation $\mathcal{T}_\gamma$ to be applied to the source image $I$. Dalca \textit{et al.} \cite{dalca2018anatomical} extended this idea to use a second image as target to register to and to incorporate an image-based loss. We follow this approach, but integrate an additional \emph{Euler2Affine} layer, which restricts the rotation component of the matrix $M$ to have a determinant of 1 to simultaneously enable a valid inverse transformation.

\subsubsection{Localisation network}
In our case, the localisation network function $f_{\textrm{loc}}()$ takes the form of a CNN encoder which learns features from the axial input image. Each of the four downsampling blocks within the encoder consists of a $3\times3\times3$ convolution layer, a dropout layer, a $3\times3\times3$ convolution layer and a $2\times2\times2$ max pooling layer. All convolution layers use zero padding and the ELU activation function. The dropout rate at the shallow layers is set to 0.3, this rate increased linearly up to 0.5 in the last layer. We forwarded the $batchsize \times 5 \times 14 \times 14 \times 256$ encoding from our localisation network through a global average pooling layer and two dense layers with 256 and 6 outputs, respectively. The latter outputs are the three rotation parameters $\phi, \theta, \psi$ and three translation parameters $\vec{t} = [t_{x}, t_{y}, t_{z}]$, which are used in the subsequent \emph{Euler2Affine} layer to convert the transformation parameters into an affine matrix representation. The transformation parameters were initialised with zero mean and a standard deviation of $1\times10^{-10}$.

\subsubsection{Euler2Affine layer}
Different to the spatial transformer layer implemented by Dalca \textit{et al.} \cite{dalca2018anatomical} our network does not learn the affine matrix directly, since we know that the transform is rigid. We restrict our transform $\mathcal{T}_\gamma$ to allow only rotations $R$, and translations, $T$. Therefore, the affine matrix in homogeneous coordinates is decomposed into two separate matrices: $M = RT$. Additionally, $R$ must fulfil the property of $|R| = 1$, which is not the case in the current implementation from \cite{dalca2018anatomical}.

\label{sec:introduction2}
We introduce a novel \emph{Euler2Affine} layer, which converts the previously estimated three Euler rotation angles $\phi, \theta, \psi$ and three translation parameters $\vec{t} = [t_{x}, t_{y}, t_{z}]$ into an affine matrix representation. 

We use the \textit{x-convention}\cite{goldstein1980euler} and interpret the coordinate transformation passive, which is the standard used in most engineering disciplines. The Euler angles are converted into a rotation matrix $R_{x}, R_{y}$ and $R_{z}$, one for each axis, 

    \begin{align}
    \label{eq:euler2affine}
    R_x =
    \begin{bmatrix}
    1 & 0 & 0 & 0\\
    0 & \cos \phi & -\sin \phi & 0\\
    0 & \sin \phi & \cos \phi & 0\\
    0 & 0 & 0 & 1 \\
    \end{bmatrix}
    \end{align}

    \begin{equation}
    R_y = 
    \begin{bmatrix}
    \cos \theta & 0 & \sin \theta & 0\\
    0 & 1 & 0 & 0\\
    -\sin \theta & 0 & \cos \theta & 0\\
    0 & 0 & 0 & 1 \\
    \end{bmatrix}
    \end{equation}
    
    \begin{equation}
    R_z = 
    \begin{bmatrix}
    \cos\psi & -\sin \psi & 0 & 0\\
    \sin \psi & \cos \psi & 0 & 0\\
    0 & 0 & 1 & 0\\
    0 & 0 & 0 & 1 \\
    \end{bmatrix}
    \end{equation}
and the final rotation component $R$ of the affine matrix is obtained by $R = R_{x} R_{y} R_{z}$ and the transformation component by 
    \begin{equation}
    T = 
    \begin{bmatrix}
    1 & 0 & 0 & t_{x}\\
    0 & 1 & 0 & t_{y}\\
    0 & 0 & 1 & t_{z}\\
    0 & 0 & 0 & 1 \\
    \end{bmatrix}.
    \end{equation}
    
$M$ is finally fed into the spatial transformer layer from Dalca \textit{et al.} \cite{dalca2018anatomical}, which applies the transformation relative to the center of image $I$. Note that their layer per default adds the identity to the affine parameters to start with the identity transform. This part was skipped. 

\subsubsection{Grid generation and sampling}
The output voxels are defined to lie on a regular target grid $G=\{G_{i}\}$, $G_{i}= (x^{t}_{i},y^t_{i},z^t_{i})$, which is sampled from the source coordinates $(x^s_{i}, y^s_{i}, z^s_{i})$ \cite{STN_Nips},
\begin{equation}
\left( \begin{array}{c} x^{s}_{i} \\\ y^{s}_{i} \\\ z^{s}_{i} \\\ 1 \end{array}\right) = \mathcal{T}_{\gamma(G_{i})} 
= M 
\left( \begin{array}{c} x^t_{i} \\\ y^t_{i} \\\ z^t_{i} \\\ 1  \end{array}\right).
\end{equation}
Linear interpolation is applied. Note that height and width coordinates were normalised such that 
$-1 \leq x^t_{i} \leq 1$ when within the spatial bounds of the target, and $-1 \leq x^s_{i} \leq 1$ when within the spatial bounds of the source image (and similarly for the $y$ and $z$ dimensions). 

\subsubsection{Masked MSE Loss}
Finally, the mean squared error (MSE) loss is computed over the image $I$, using the ground truth  ($\hat\gamma$) and the predicted transformation parameters ($\gamma$),
\begin{equation}
    Loss_{MSE} = \dfrac{1}{2}\left(\left(\mathcal{T}(I,RT) - \mathcal{T}(I, \hat R \hat T)\right) \cdot V_{\hat I}\right)^2 
\label{eq:MSE}
\end{equation}

After the transformation $\mathcal{T}(I, \hat{R} \hat{T})$, the resulting image will contain larger areas filled with zeros at the border. We therefore apply a mask $V_{\hat{I}}$ to omit these regions from the loss.

\begin{figure*}[t]
\centering
  \includegraphics[width=\linewidth]{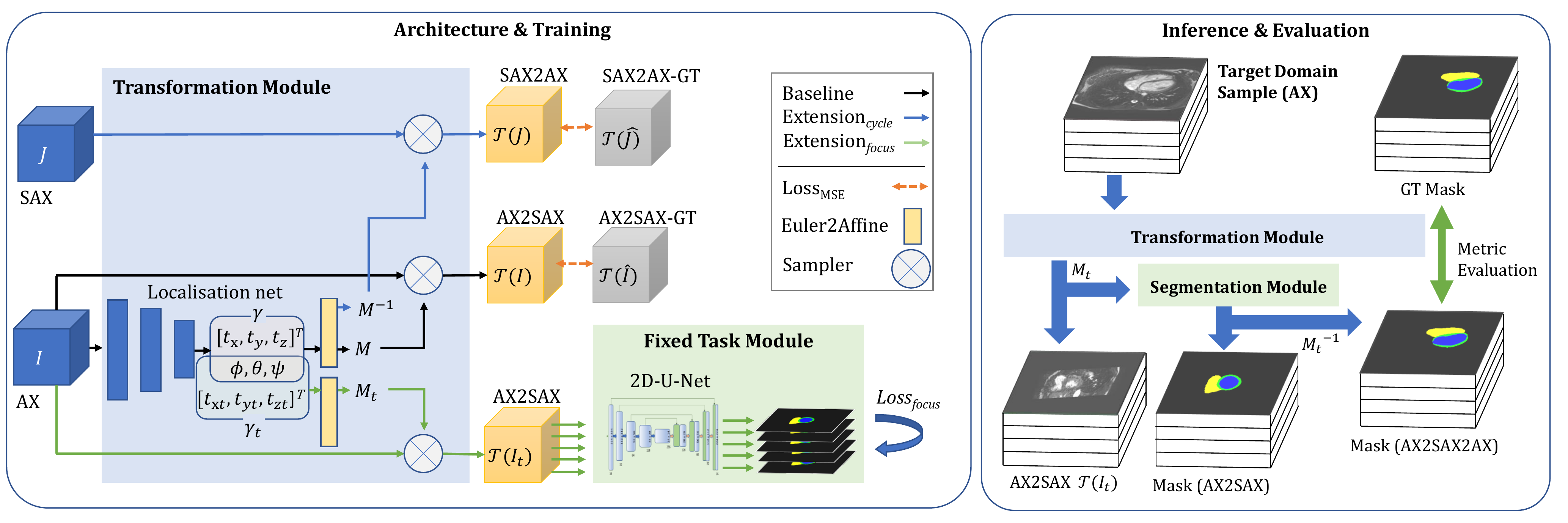}
  \caption{The proposed network with both extensions during the training and final evaluation process. During training, AX and SAX CMR image pairs ($I$ and $J$) and a pre-trained task network are required to learn the transformation parameters $\gamma$ and $\gamma_{t}$. During inference, only an AX CMR stack is used. The model automatically transforms this volume into a SAX orientation and performs the task, e.g. a segmentation, and this volume is then transformed back into the original domain by the inverse transformation $M^{-1}_{t}$.}
  \label{fig:pipeline}
\end{figure*}

\subsection{Transformation module (extension)}
\label{sec:trans_extension_method}

When the baseline transformation model is learned (cf. black line in Fig. \ref{fig:pipeline}) and the segmentation module is applied on the transformed image $\mathcal{T}(I)$, results were not satisfactory and several problems could be identified:

\begin{itemize}
    \item Learning of the localisation network is slow and unstable. Depending on the weight initialisation, some training runs did not converge at all or they reached only semi-optimal generalisation capabilities. 
    The solution to these issues is to additionally compute the  inverse transformation in the \emph{Euler2Affine} layer and apply it to the opposite SAX to AX transformation. This adds a second MSE-loss component into the loss, which we refer to as \emph{cycle consistency} (details in \S \ref{sec:cycle}).
    \item Even when the baseline model converged, it was faced with another problem. Most resulting volumes barely covered the ventricles close to the apex, i.e. most of the predicted transformations included a very plausible rotation in terms of how a short axis stack usually looks like, but relevant lower slices of the volume were cut off. This is due to the fact that the translation required for the task is not enough enforced in the baseline design (Eqn. \ref{eq:MSE}). 
    We solved this issue by branching out after the localisation network into a second sub-network, which has the freedom of learning independent translation parameters $[t_{xt}, t_{yt}, t_{zt}]^{T}$, but back-propagate the gradients into the same localisation network  
     (details in \S \ref{sec:prop}). Furthermore, in-plane weighting on the masked MSE loss helped to weight errors in the center of the grid region higher compared to areas close to the border (details in \S \ref{sec:cycle}).
\end{itemize}
All in all, the correct rotational parameters are mostly enforced by the cycle consistency (simultaneous transformation of AX input to SAX orientation and SAX input to AX orientation), while the translation parameters are fine-tuned by the second branch, which shifts the image information on the target grid.

\subsubsection{Cycle consistency}
\label{sec:cycle}
To improve the general transformation stability, we incorporate a cycle consistency (cf. blue line in Fig. \ref{fig:pipeline}). A similar approach was followed by \cite{Kim2019} for deformable image registration using a flow field. However, we do not incorporate a second encoder for learning of a second transformation, but only apply the inverted matrix $M^{-1}$ from our single encoder to the second input $J$, which is the SAX volume. In particular, our \emph{Euler2Affine} layer always delivers two outputs: the forward transformation and the corresponding inverted backward transformation, while the transformation parameters $\gamma$ are optimised by an image-based loss. 
The final masked cycle loss is composed of two MSE losses and is defined as
\begin{align}
    Loss_{Cycle} &= \dfrac{1}{2}\left(\left(\mathcal{T}(I,RT) - \mathcal{T}(I, \hat R \hat T)\right)\cdot V_{\hat I} \cdot W \right)^2  \nonumber \\
    &+ \dfrac{1}{2}\left(\left(\mathcal{T}(J,R^{-1}T^{-1}) - \mathcal{T}(J, \hat R^{-1} \hat T^{-1})\right)\cdot V_{\hat J} \cdot W\right)^2 .
    \label{eqn:cycle}
\end{align}
with $V_{\hat{I}}$ and $V_{\hat{J}}$ being the respective masks (cf. Eqn. \ref{eq:MSE}). $W$ is a linear in-plane image gradient per slice interpolated between one in the center and zero at the border (cf. Fig. 2 in the Supplemental Material).

\subsubsection{Task-oriented guidance with probability maps}
\label{sec:prop}

The goal is to weakly guide the transformation module without the necessity of feeding a task-specific ground truth (in our case the masks) as input, which makes the approach more task-independent and existing task models can be incorporated more easily. 
We injected a pre-trained 2D U-Net (cf. \S \ref{sec:unet}) with fixed weights. This model returns the pixel-wise class probabilities for the LV, RV, MYO and background class during the forward pass and during the back-propagation step, we maximize the number of voxels for the foreground classes with a probability higher than a given threshold $r$, with $r=0.9$. We set this threshold very close to 1, as we are only interested in very reliable predictions. The additional loss is defined as 

\begin{equation}
Loss_{Focus} = 1 - \dfrac{ \sum_{j=1}^C\sum_{i=1}^N 1_{q_{ij} > r}(q_{ij})}{CN}.
\label{eq:lossfocus}
\end{equation}
Simultaneously, we increased the number of outputs of the localisation net from 6 to 9 parameters: The three additional parameters are a second set of translation parameters $\vec{t_{t}}$ and form with the rotation $R$ another rigid transformation matrix $M_{t}$, that is additionally applied to $I$, yielding $I_{t}=\mathcal{T}(I,RT_{t})$ (cf. green line in Fig. \ref{fig:pipeline}).
$I_{t}$ can be understood as the domain adapted image that is fed into the segmentation module.

The gradients are back-propagated into the same localisation network.
This allows this network branch to control a shift to maximize the number of predicted foreground voxels and to therefore reduce cutting of relevant structures due to sub-optimal translation. 

\subsubsection{Summary}
Fig. \ref{fig:pipeline} summarizes the introduced pipeline with both major extensions. Fig. \ref{fig:inference_1} and Fig. \ref{fig:inference_2} show the intermediate steps for the transformation module and the injected task module. The extended transformer module consists of a localisation encoder which indirectly regresses on the three Euler angles $\phi, \theta, \psi$ and two sets of translation parameters $\vec{t}$ and $\vec{t_{t}}$ by using image-based losses. Afterwards, it branches into two sub-networks, both simultaneously optimize the rotation and their corresponding set of translation parameters. This enforces the output volume to include relevant anatomical regions of interest  (e.g. not shifting the target structure out of the 3D volume). The transformation module does neither require a ground truth mask for the actual task nor the target transformation parameters. The gradients are determined by the combination of two MSE losses  (cf. Eqn. \ref{eqn:cycle}) and a focus loss (cf. Eqn. \ref{eq:lossfocus}).
The overall loss can be written as 
\begin{equation}
    Loss = \alpha_{1} Loss_{Cycle} + \alpha_{2} Loss_{Focus}, 
    \label{sec:finalLoss}
\end{equation}
where $\alpha_{1}= 1.0$, and $\alpha_{2}= 0.1$. The sensitivity of these parameters is low as long as $\alpha_{1} > \alpha_{2}$. Pseudo-code is provided in the Supplemental Material.

\section{Experiments}
\subsection{Datasets}
\subsubsection{Tetralogy of Fallot dataset}
\label{sec:tof}
\begin{figure}
\centering
  \includegraphics[width=\linewidth]{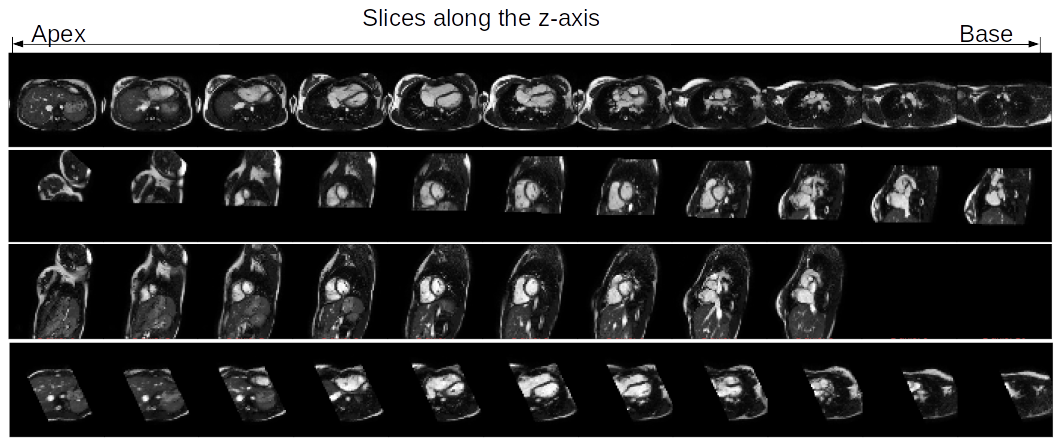}
  \caption{One example stack of an AX, AX2SAX ground truth, SAX and SAX2AX ground truth image stack (from top to bottom).}
  \label{fig:ax_sax_example_pair}
\end{figure}

A multi-centric heterogeneous cine-SSFPs CMR TOF data set from the German Competence Network for Congenital Heart Defects was used (study identifier: NCT00266188, title: \textit{Non-invasive Imaging and Exercise Tolerance Tests in  Post-repair Tetralogy of Fallot - Intervention and Course in Patients Over 8 Years Old}). This TOF dataset constitutes one of the largest compiled data set of this pathology to date. The data was acquired at 14 different sites between 2005-2008 on 1.5T and 3T machines; further descriptions can be found in \cite{Sarikouch2011}. Two disjunct  sub-cohorts were defined: 
\begin{itemize}
    \item TOF sub-cohort 1: 193 patients were used to separately fine-tune a U-Net on short-axis images. The images were manually contoured on five phases, including ES and ED, by an experienced pediatric cardiologist. The CMR volumes of this sub-cohort have an average resolution of $238.52\pm{62.95} \times 248.79\pm{49.74} \times 13.93\pm{3.20}$ and a spacing of $1.38\pm{0.19} \times 1.38\pm{0.19} \times 8.02\pm{1.31}$ mm$^3$ (X/Y/Z).
     \item TOF sub-cohort 2: 81 patients had both, axial and short-axis acquisitions and were used for training and evaluating of the proposed pipeline. An example is shown in Fig. \ref{fig:ax_sax_example_pair}. The images were manually contoured on ES and ED phase by the same cardiologist. The original transformation was obtained from the DICOM tags, which specify the image position (tag: 0020,0032) and the image orientation (tag: 0020,0037) in world coordinate space. The average resolution of the AX CMR is $260.21\pm{45.10} \times 257.12\pm{51.67} \times 22.01\pm{5.60}$ with a spacing of $1.30\pm{0.1} \times 1.30\pm{0.1} \times 6.02\pm{1.0}$ mm (X/Y/Z). The SAX CMR images have a resolution of $259\pm{50} \times 261\pm{46} \times 14.\pm{3.32}$ and a spacing of $1.35\pm{0.11} \times 1.35\pm{0.11} \times 8.0\pm{1.02}$ mm$^3$. 
\end{itemize} 

\subsubsection{ACDC Dataset}
This dataset was introduced as part of the \textit{Automatic Cardiac Diagnosis Challenge} (ACDC) \cite{Bernard2018a} and covers adults with normal cardiac anatomy and the following four cardiac pathologies: systolic heart failure with infarction, dilated cardiomyopathy, hypertrophic cardiomyopathy and abnormal right ventricular volume. Each pathology is represented by 20 patients with labels for the RV, LV and MYO at ES and ED phase. A detailed description of the data set properties can be found in \cite{Bernard2018a}.

\subsection{Pre-processing}

\subsubsection{TOF sub-cohort 1 and ACDC}
\label{sec:subcohort1}
For training of the U-Net, a composition of different augmentation methods from the library Albumentations \cite{albumentations} was applied to the image slices with a probability of 80\%: random rotation of 90$^\circ$, shifts with a shift factor range of $0.025$, scaling with a scale factor range of $0.1$, brightness and contrast manipulations with a random contrast limit of $0.05$ and random grid distortion. 

All images and masks were center cropped. Bi-linear interpolation and anti-aliasing for the CMR images and nearest neighbor interpolation for the masks was used. Grey values were clipped on the $0.999$ quantile and min/max-normalised.

\subsubsection{TOF sub-cohort 2}
\label{sec:subcohort2}
The same pre-processing methods, adopted to the 3D space, were applied to this sub-cohort except for the augmentation part, which was skipped to avoid implausible deformations between the two 3D volumes. Resizing was also skipped, instead, zero-padding was performed to keep the original aspect ratio. A source and target grid size of $ 224 \times 224 \times 96$ was defined.
Besides that, all volumes were re-sampled to a uniform isotropic voxel size of $1.5$~mm$^3$ with linear interpolation. Additionally, the AX images were transformed to the SAX orientation by resampling with the direction and origin of the corresponding SAX. The SAX origin was further shifted by 10~mm along the $z$-axis and the size 
was increased by 10 mm; therefore the physical size of the $AX2SAX$ volumes along $z$ increased by 20 mm. This extends the target grid (here: target with respect to the transformation, not the domain adaptation) and decreases the possibility of cropping the ventricles at the base or apex. When resampling the multi-label masks during AX to SAX transformation, several labels could fall into the same voxel, violating the assumption that our anatomical labels are disjunct for each location. To avoid this, we first scaled the labels by 100 and then interpolated linearly. Then, an argmax operation was applied to every voxel to choose the label with the highest value.

In seven cases, the image pairs had to be registered beforehand due to patient movement between the acquisitions. The iterative closest point algorithm \cite{besl_method_1992} was used to estimate the rigid transformation between the reconstructed surfaces of LV and RV. 
The estimated  orientation and translation were applied to the axial image. Independently to that, note that slight temporal misalignment is evident in the selected phases in many AX-SAX pairs. 

\begin{figure}
\centering
  \includegraphics[width=0.8\linewidth]{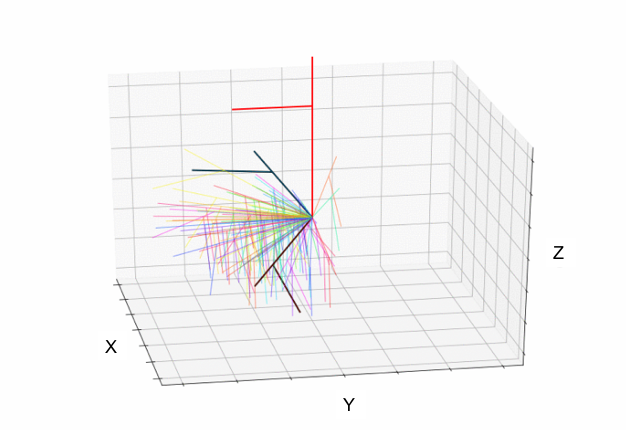}
  \caption{Relative orientation between AX and SAX captured in our data, here shown symbolically by applying this transformation to a dummy geometry (red flag). Note that the relative translation is omitted.}
  \label{fig:saxvariations}
\end{figure}

\subsection{Training and Evaluation}
\subsubsection{Segmentation module}
\label{sec:segmentationmoduleval}
Previous work of the authors on the same overall cohort addressed short-axis segmentation using a 2D U-Net approach trained on ACDC and TOF data \cite{Koehler2020}. In this work, different splits were used. 
The TOF sub-cohort 1 and the ACDC data were shuffled and split into a training (75\%) and validation (25\%) set to monitor the training progress and to avoid overfitting. The ACDC dataset was split with respect to the four pathologies and the healthy subgroup in a stratified manner; 75 patients were used in the training split (15 per pathology) with 1426 slices and 25 patients in the validation split (5 per pathology) with 476 slices.
Splitting of the TOF data resulted in 144 patients with 9975 slices for the training split and 39 patients with 3355 slices for validation.

The network was trained with a batch size of 32. Early stopping with a patience of 20 epochs without change in the validation loss was applied. The initial learning rate was set to $1\times10^{-3}$; this learning rate was decreased by a factor of 0.3 (minimal learning rate: $1\times10^{-8}$) after ten epochs without any gain in the loss. The standard Adam optimizer was used. The segmentation model had 19.4M trainable parameters.
All training runs were performed in parallel on two TITAN RTX graphics cards with a distributed mirror strategy. 

\subsubsection{Transformation module}

The transformation module was trained in a four-fold cross-validation manner. For this, the data from the 81 TOF patients of TOF sub-cohort 2 were shuffled and split. Each training fold represents about 61 patients (61 ED and 61 ES AX-SAX 3D CMR image pairs), the corresponding validation split contains about 20 patients (20 ED and 20 ES AX-SAX 3D CMR image pairs).

We applied early stopping with a patience of 10 epochs without change on the validation loss to avoid overfitting. The initial learning rate was set to $1\times10^{-3}$, this learning rate was decreased by a factor of 0.3 (minimal learning rate: $1\times10^{-8}$) after five epochs without any gain in the loss. The standard Adam optimizer was used. A batch size of two was chosen. Most of the models converged after 30-50 epochs. The transformation module had 23M parameters, where 3.6M parameters were trainable and 19.4M were fixed and belonged to the injected task network.

\begin{figure}
\centering
  \includegraphics[width=\linewidth]{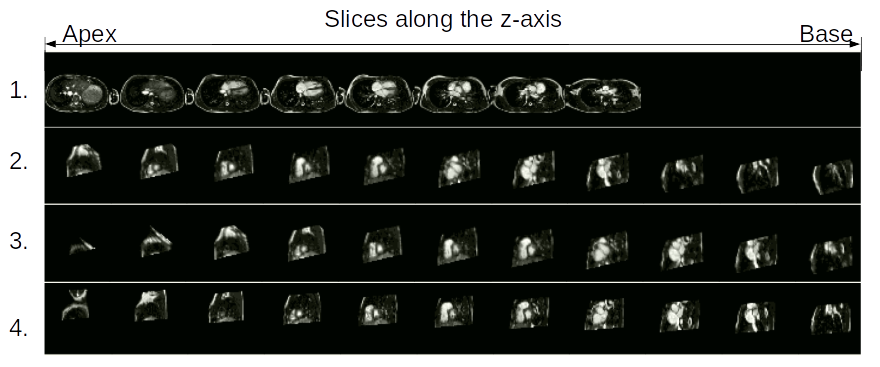}
  \caption{Example slices in z-directions through the 3D volumes comparing an  
  1) Original AX, 2) Predicted AX2SAX CMR using $M$, 3) Predicted AX2SAX shifted by the second set of translation parameters $\vec{t_t}$. 4) Ground truth AX2SAX.}
  \label{fig:inference_1}
\end{figure}

\begin{figure}
\centering
  \includegraphics[width=\linewidth]{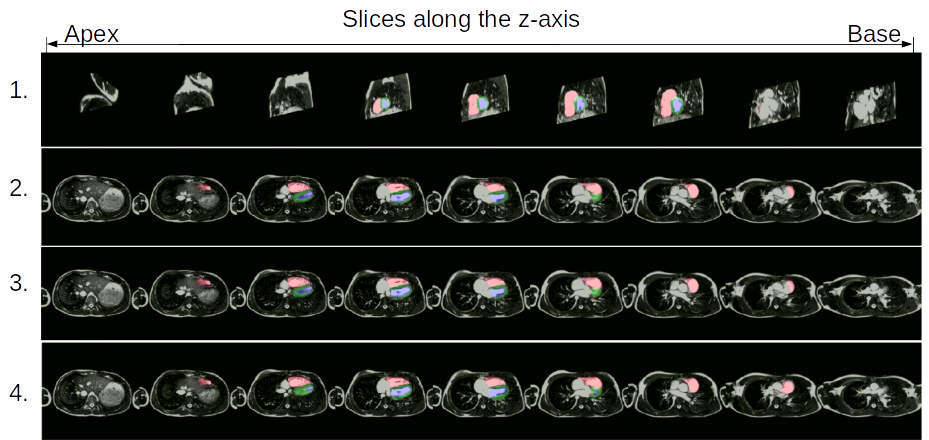}
  \caption{Predicted segmentation on the transformed volumes. 1) On the AX2SAX transformed with $M_t$. 2) Predicted SAX contours back transformed with the inverse matrix $M_t^{-1}$ and overlayed over the input AX CMR stack. 3) Post-processed AX mask 4) GT mask on the input AX CMR stack.}
  \label{fig:inference_2}
\end{figure}

\subsubsection{Measure of Performance}
For metric computation, we used the library Medpy. The performance of the segmentation module and the full pipeline are assessed by the 3D Dice similarity index between the ground truth and predicted mask, $I_{T}$ and $I_{P}$, as $D= 2(|I_{T}\cap I_{P}|)/(|I_{T}|+|I_{P}|)$. 
Furthermore, we report the 3D Hausdorff surface distance ($d_{H}$), which computes the maximum symmetric distance between $I_{T}$ and $I_{P}$. In cases where our model was not able to predict any mask, which occurred mainly in the baseline model (cf. Section \ref{sec:trans_base}) due to implausible transformations, we \textit{excluded} this volume from the $d_{H}$ computation. 
Furthermore, the difference between ground truth volume and predicted volume of each foreground region is provided as a clinical measure. Please note that we did not perform a comparison of the predicted transformation to the original transformation parameters, as we regard them as a weak ground truth (e.g. a slight shift in the translation parameters might equally well enable the task-network to generate meaningful or even better output).

\section{Results}
\label{sec:results}
This section first provides more visual insight into the data and then describes the results for our unsupervised domain adaptation method, which enables to segment axial CMR images without the need of any axial ground truth during training. The segmentation module was assessed by applying it to different independent data sets (Table \ref{table:dice_scores_unet}). Subsequently, results for the components of the full pipeline and differences to the baseline transformation module were systematically evaluated (Table \ref{table:dice_scores_per_experiment}).

\subsection{Visual Analysis}
Each patient has an individual orientation of the heart. Therefore, it is common to capture CMR images in SAX view, as the obtained slices have a direct relation to the anatomy. Axial images are oriented according to the thorax orientation. A rigid relative transformation between both images varies heavily from patient to patient, and, besides that, captures different physical volumes. Fig. \ref{fig:saxvariations} shows the relative rotations of the SAX images in our data cohort with respect to a dummy geometry. The high variation makes it impossible to simply just apply the average rotation to all AX images in order to close the domain gap.

\begin{table*}[t]
\centering
\caption{3D DICE and Hausdorff distance (in mm) results for the same U-Net module applied to different source (SAX) and target domain (AX) samples.}
\begin{resizebox}{\linewidth}{!}{
\begin{tabular}{l|c|cc|cc|cc|cc|cc|cc}
\\
&  
&\multicolumn{4}{c}{LV} \vline
&\multicolumn{4}{c}{MYO}
&\multicolumn{4}{c}{RV}

\\

Experiment
&
&\multicolumn{2}{c}{mean$\pm{SD}$}\vline
&\multicolumn{2}{c}{median}\vline
&\multicolumn{2}{c}{mean$\pm{SD}$}\vline
&\multicolumn{2}{c}{median}\vline
&\multicolumn{2}{c}{mean$\pm{SD}$}\vline
&\multicolumn{2}{c}{median}
\\

& GT$^1$
& D $\uparrow$
& $d_{H}$ $\downarrow$
& D $\uparrow$
& $d_{H}$ $\downarrow$
& D $\uparrow$
& $d_{H}$ $\downarrow$
& D $\uparrow$
& $d_{H}$ $\downarrow$
& D $\uparrow$
& $d_{H}$ $\downarrow$
& D $\uparrow$
& $d_{H}$ $\downarrow$
\\
\hline

ACDC val set (SAX)      
& SAX 
& 0.93$\pm{0.04}$
& 5.38$\pm{4.33}$
& 0.95
& 3.87
& 0.87$\pm{0.03}$
& 6.58$\pm{4.23}$
& 0.87
& 4.79
& 0.85$\pm{0.09}$
& 12.79$\pm{5.89}$
& 0.87
& 11.29

\\
\hline
TOF sub-cohort 1 val set (SAX)  
& SAX 
& 0.93$\pm{0.03}$
& 6.24$\pm{5.91}$
& 0.93
& 5.10
& 0.81$\pm{0.04}$
& 9.38$\pm{4.85}$
& 0.82
& 8.37
& 0.90$\pm{0.04}$
& 11.33$\pm{7.20}$
& 0.91
& 9.49

\\
\hline
TOF sub-cohort 2 (AX) *
& AX 
& 0.78$\pm{0.17}$
& 15.24$\pm{9.85}$
& 0.83
& 11.55
& 0.63$\pm{0.14}$
& 19.95$\pm{8.63}$
& 0.67
& 17.03
& 0.52$\pm{0.19}$
& 33.98$\pm{8.30}$
& 0.56
& 33.32

\\
\hline
TOF sub-cohort 2 (AX2SAX $\mathcal{T}(I,\hat{R}\hat{T})$)  
& AX 
& 0.82$\pm{0.14}$
& 11.70$\pm{6.99}$
& 0.86
& 9.46
& 0.61$\pm{0.13}$
& 17.02$\pm{7.55}$
& 0.65
& 14.54
& 0.75$\pm{0.15}$
& 17.87$\pm{6.77}$
& 0.79
& 16.34
\\
\hline
\multicolumn{14}{p{22cm}}{ $^1$ Mask ground truth used to compute the metric.
AX=axial, SAX=short-axis,  LV=left ventricle, MYO=myocardium, RV=right ventricle, D=Dice index, $d_{H}$=Hausdorff distance, ACDC=Automated Cardiac Diagnosis Challenge, TOF=tetralogy of fallot, GT=ground truth, SD=standard deviation}\\
\multicolumn{14}{p{22cm}}{ * = Domain Gap}
\end{tabular}
}\end{resizebox}
\label{table:dice_scores_unet}
\end{table*}

\begin{table*}[t]
\centering
\caption{3D Dice and Hausdorff distance (in mm) results for the components of our domain adaptation approach after four-fold cross validation on TOF sub-cohort 2.}
\begin{resizebox}{\linewidth}{!}{
\begin{tabular}{l|c|cc|cc|cc|cc|cc|cc}
\\
&  
&\multicolumn{4}{c}{LV} \vline
&\multicolumn{4}{c}{MYO}
&\multicolumn{4}{c}{RV}

\\

Experiment
&
&\multicolumn{2}{c}{mean$\pm{SD}$}\vline
&\multicolumn{2}{c}{median}\vline
&\multicolumn{2}{c}{mean$\pm{SD}$}\vline
&\multicolumn{2}{c}{median}\vline
&\multicolumn{2}{c}{mean$\pm{SD}$}\vline
&\multicolumn{2}{c}{median}
\\

& GT$^1$
& D$\uparrow$
& $d_{H}$ $\downarrow$
& D$\uparrow$
& $d_{H}$ $\downarrow$
& D$\uparrow$
& $d_{H}$ $\downarrow$
& D$\uparrow$
& $d_{H}$ $\downarrow$
& D$\uparrow$
& $d_{H}$ $\downarrow$
& D$\uparrow$
& $d_{H}$ $\downarrow$
\\
\hline

Baseline $Loss_{MSE}$ w/o ip *
& AX 
& 0.62$\pm{0.37}$
& 11.02$\pm{6.08}$
& 0.83
& 8.94
& 0.47$\pm{0.29}$
& 14.79$\pm{5.21}$
& 0.62
& 13.04
& 0.55$\pm{0.34}$
& 19.31$\pm{7.54}$
& 0.72
& 18.52
\\
\hline

Baseline $Loss_{MSE}$ w/ ip *
& AX 
& 0.61$\pm{0.38}$
& 12.17$\pm{6.91}$
& 0.82
& 10.05
& 0.46$\pm{0.29}$
& 16.68$\pm{10.21}$
& 0.61
& 14.35
& 0.52$\pm{0.34}$
& 21.10$\pm{7.43}$
& 0.69
& 19.76
\\
\hline

$Loss_{MSE}$ w/o ip $+ Loss_{Focus}$ 
& AX 
& 0.83$\pm{0.13}$
& 11.82$\pm{6.40}$
& 0.85
& 9.85
& 0.64$\pm{0.12}$
& 15.74$\pm{6.56}$
& 0.66
& 13.93
& 0.71$\pm{0.15}$
& 22.18$\pm{7.58}$
& 0.75
& 21.21

\\
\hline
$Loss_{MSE}$ w/ ip $+ Loss_{Focus}$  
& AX 
& 0.82$\pm{0.11}$
& 12.67$\pm{9.52}$
& 0.85
& 10.05
& \textbf{0.66}$\pm{0.10}$
& 15.68$\pm{5.97}$
& \textbf{0.67}
& 14.04
& 0.66$\pm{0.13}$
& 25.70$\pm{7.52}$
& 0.68
& 24.69

\\
\hline
$Loss_{Cycle}$ w/o ip *
& AX 
& 0.82$\pm{0.15}$
& 12.00$\pm{10.04}$
& 0.85
& 9.25
& 0.63$\pm{0.13}$
& 15.54$\pm{6.52}$
& \textbf{0.67}
& 13.38
& 0.70$\pm{0.16}$
& 22.03$\pm{8.04}$
& 0.75
& 20.49

\\
\hline
$Loss_{Cycle}$ w/ ip *
& AX 
& 0.81$\pm{0.16}$
& 11.33$\pm{6.00}$
& 0.85
& 9.75
& 0.61$\pm{0.14}$
& 15.60$\pm{6.44}$
& 0.65
& 13.93
& 0.73$\pm{0.14}$
& 19.16$\pm{7.50}$
& 0.77
& 17.80

\\
\hline
Full pipeline $Loss$ w/o ip
& AX 
& 0.83$\pm{0.12}$
& 11.03$\pm{6.06}$
& \textbf{0.86}
& 9.08
& 0.63$\pm{0.12}$
& 19.95$\pm{8.63}$
& 0.65
& 17.03
& 0.73$\pm{0.14}$
& 20.41$\pm{7.31}$
& 0.77
& 19.72

\\
\hline
Full pipeline $Loss$ w/ ip
& AX 
& \textbf{0.86}$\pm{0.06}$
& \textbf{9.62}$\pm{3.97}$
& \textbf{0.86}
& \textbf{8.34}
& 0.65$\pm{0.08}$
& \textbf{13.72}$\pm{3.70}$
& 0.66
& \textbf{12.85}
& \textbf{0.77}$\pm{0.10}$
& \textbf{18.31}$\pm{5.77}$
& \textbf{0.79}
& \textbf{17.92}

\\
\hline

\multicolumn{14}{p{22cm}}{ $^1$ Mask ground truth used to compute the metric.
AX$=$axial, SAX$=$short-axis,  LV$=$left ventricle, MYO$=$myocardium, RV$=$right ventricle, D$=$Dice index, $d_{H}=$Hausdorff distance, GT$=$ground truth, SD$=$standard deviation, ip$=$inplane-weighting}\\
\multicolumn{14}{p{22cm}}{$*$ $=$ Baseline and $Loss_{cycle}$ models predicted empty masks for 41 and 2 transformed CMR images, respectively. Those patients were excluded in the $d_{H}$ metric computations.}
\end{tabular}
}\end{resizebox}
\label{table:dice_scores_per_experiment}
\end{table*}

\subsection{Segmentation Module}
Table \ref{tab:unetlossalpha} shows the results of the sensitivity analysis for the loss weighting of the segmentation module. Using only the $CE$ or the $SDL$ performed always worse than combing both losses. Reducing the influence of $CE$ to a half ($w_{seg} = 0.5$) improved the training and validation scores by $0.5$\% $DSC$ compared to a combined loss without weighting; and even improved by 3\% $DSC$ compared to only using the $BCE$ ($w_{seg} = 0$).

\begin{table}[ht]
    \centering
    \caption{Sensitivity of the loss weighting parameter $w_{seg}$ within the training of our segmentation module. The best weighting parameter is marked in bold. The $DSC$ score of this table represents the average over the three foreground classes.}
    \begin{tabular}{c|cc}
         Weighting of $w_{seg}$&  Training $DSC$ & Validation $DSC$\\
         \hline
         Only BCE& 0.8604 & 0.8527 \\
         0.2 &  0.8922 & 0.8840 \\
         \textbf{0.5} & \textbf{0.8961} & \textbf{0.8882}\\
         0.7& 0.8936 & 0.8852 \\
         1.0 & 0.8917 & 0.8832\\
        Only SDL & 0.8916 & 0.8835\\ 
    \end{tabular}
    \label{tab:unetlossalpha}
\end{table}

The model achieved a mean$\pm{}$stddev Dice $DSC$ and Hausdorff distance $d_{H}$ (in mm) of 
$0.93 \pm 0.04$ / $5.38 \pm 4.33$ for LV, 
$0.87 \pm 0.03$ / $6.58 \pm 4.23$ for MYO and 
$0.85 \pm 0.09$ / $12.79 \pm 5.89$ for RV 
on the ACDC SAX validation split and 
$0.93 \pm 0.03$ / $6.24 \pm 5.91$ for LV, 
$0.81 \pm 0.04$ / $9.38 \pm 4.85$ for MYO and 
$0.90 \pm 0.04$ / $11.33 \pm 7.20$ for RV for the TOF sub-cohort 1 validation split (SAX).

Thus, the Dice score of the module was $5 \%$ better for RV segmentation on the TOF data in comparison to the ACDC data, as the training set included a considerable amount of slices covering this pathology. 
Simultaneously, the Dice was $6 \%$ worse for MYO segmentation of TOF patients compared to the ACDC dataset which aligns with the previous investigations on both datasets from Koehler \textit{et al.} \cite{Koehler2020}.

Fig. \ref{fig:dice_baseline} summarizes the results. The model performances are in the range of the state-of-the-art considering the ACDC challenge \cite{Bernard2018a}, however, note that metrics are computed on a different subset of patient data than in \cite{Bernard2018a}, since ground truth contours for the test data set are not openly available.

Throughout the experiments, morphological closing and a label-wise filtering of the biggest connected component improved the final mean Dice per label by $1-2\%$. All presented results include these post-processing steps. 

\subsection{Domain Gap}
\begin{figure}
\centering
  \includegraphics[width=\linewidth]{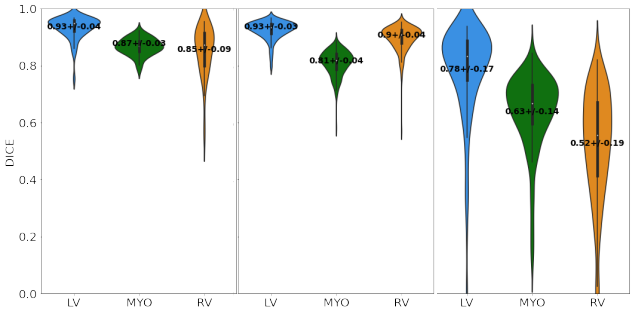}
  \caption{Violin plots showing the performance (mean$\pm SD$ 3D Dice) of the segmentation module. From left to right: 1) On the ACDC SAX validation split, 2) On the TOF SAX sub-cohort 1 validation split, 3) On the TOF AX sub-cohort 2. The latter shows the domain gap of a model trained on SAX samples but applied to axial data. Corresponding results in Tab. \ref{table:dice_scores_unet}, 1st, 2nd and 3rd row.}
  \label{fig:dice_baseline}
\end{figure}

Applying the SAX model on axial slices without domain adaptation shows a huge performance decrease, as depicted on the right side of
Fig. \ref{fig:dice_baseline} and in Fig. \ref{fig:domain_gap}. 

\begin{figure}
\centering
  \includegraphics[width=0.8\linewidth]{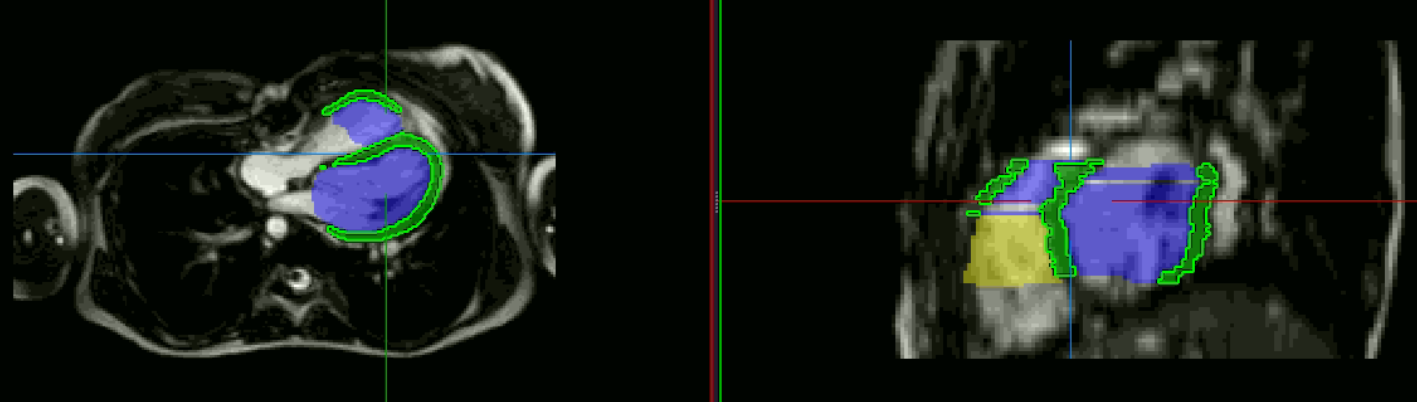}
  \caption{Axial (left) and sagittal (right) view of an implausible U-Net prediction (trained only on SAX images) which was applied on a stack of AX images without incorporating the domain adaptation approach. MYO in green, LV in blue, RV in yellow.}
  \label{fig:unet_on_ax}
\end{figure}

The mean Dice / $d_{H}$ on the never seen TOF AX slices from TOF sub-cohort 2 dropped to 
$0.78 \pm {0.17}$ / $15.24 \pm {9.85}$ for LV,
$0.63 \pm {0.14}$ / $19.95 \pm {8.63}$ for MYO, and 
$0.52 \pm {0.19}$ / $33.98 \pm {8.30}$ for RV.
Especially the mean $d_{H}$ indicates a large distribution shift between the two domains.
An example prediction is given in Fig. \ref{fig:unet_on_ax}, which shows impl05ausible contours.

\subsection{Upper limit with respect to a direct regression on the transformation parameters}
\label{sec:upper_limit}
We can define an upper bound of the performance that can possibly be achieved by formulating the transformation problem as a direct regression approach on the transformation parameters. The upper bound is computed by applying the relative ground truth transformation between the paired data on the axial ground truth image. The major problem here is that we have to work with grids of fixed sizes, which do not necessarily accommodate the regions of interest in the source domain (SAX). 
The mean Dice / $d_{H}$ results that can be achieved after application of the U-Net on the ground truth transformed axial volume were 
$0.82 \pm {0.14}$ / $11.70 \pm {6.99}$ for LV,
$0.61 \pm {0.13}$ / $17.02 \pm {7.55}$ for MYO, and 
$0.75 \pm {0.15}$ / $17.87 \pm {6.77}$ for RV.
More details are provided in Fig. \ref{fig:unet_on_resample_ax2sax}.

\subsection{Transformation module (baseline)}
This and subsequent experiments are conducted on the TOF sub-cohort 2 data set. 
Applying the segmentation module separately after the basic transformation module (Section \ref{sec:trans_base}) produces unsatisfactory low mean Dice / Hausdorff scores of 
$0.62\pm{0.37}$ / $16.20\pm{10.49}$ for LV, 
$0.47\pm{0.29}$ / $19.15\pm{8.75}$ for MYO and 
$0.55\pm{0.34}$ / $22.44\pm{8.53}$ for RV; indicating that the localisation net needs more guidance than the $Loss_{MSE}$ (Eq. \ref{eq:MSE}) provides. Especially, we observed that some relevant CMR regions of interest were cut after transformation. Note, the baseline transformation module did not converge on every fold, which resulted in 41 3D volumes with empty predicted masks. These volumes were neither included into the $d_{H}$ metrics of Tab. \ref{table:dice_scores_per_experiment} nor into the corresponding Fig. \ref{fig:ax1sax0focus0Noinplane}, as the $d_{H}$ is not defined for empty volumes.

\subsection{Transformation module (single components)}
Tab. \ref{table:dice_scores_unet} summarizes the contribution of each single component starting from the baseline architecture. 
The integration of the cycle-consistency considerably improved the mean Dice for each label by 
$15-20 \%$, the median improved by 
$2-5\%$, depending on the label. Note that the $d_{H}$ was within the same range for the baseline and cycle-extension experiments, but the latter predicted empty masks for 160 out of 162 volumes compared to 121 predicted by the baseline.
A very similar picture can be drawn by looking at the contribution of the $Loss_{focus}$ when combining it with the basic $Loss_{MSE}$, meaning that both major proposed extensions yield a similar performance gain. The integration of the $Loss_{focus}$ led to models which predicted empty masks for all 162 3D volumes.
The in-plane weighting changed the mean Dice score of the intermediate experiments between 
$-3\%$ and 
$+3\%$. 
Most of them performed worse after this extension. Only the full pipeline was able to benefit from this extension for all labels with a mean Dice / $h_{D}$ enhancement of
$2-4\%$ / $1-6$mm.

\subsection{Transformation module (full extension)}
\label{sec:ResultFullExtension}
Fig. 1 in the Supplemental Material provides example learning curves for the full pipeline, which provide further insights why $\alpha_{1}$ and $\alpha_{2}$ in Eqn. \ref{sec:finalLoss} have been chosen at such different scales. It turned out that the model did not start to transform the input image at all if $\alpha_{2}$ was equal or bigger than $\alpha_{1}$. This is due to the fact that starting to rotate the AX input image usually leads to views which were ever seen by the segmentation module, leading to sub-optimal segmentation results. The $Loss_{Focus}$ typically gets better as the transformation yields images which do resemble the SAX view. If $\alpha_{2}$ is less than one twentieth of $\alpha_{2}$, no or only very little task dependant translation $\vec{t_{t}}$ was learnt. The  learning progress of the transformation module (one forward propagation of an example AX image stack after each training epoch) is shown for an example data set in our  supplemental video\footnote{Video of the learning progress: \url{https://github.com/Cardio-AI/3d-mri-domain-adaptation}}.

\begin{figure}
  \centering
      \includegraphics[width=.45\textwidth]{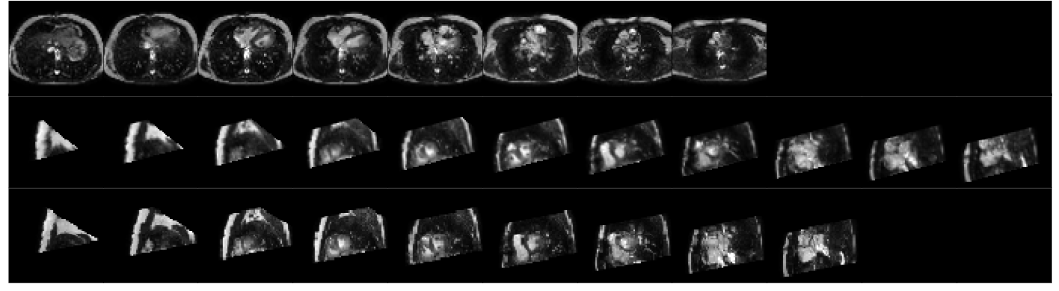}
    \hrule
      \includegraphics[width=.45\textwidth]{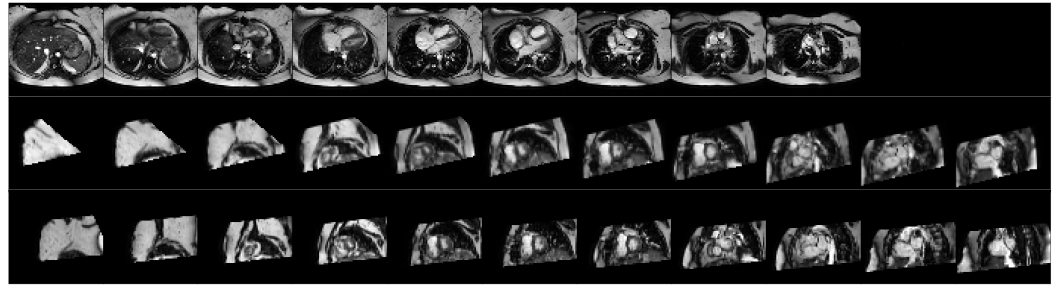}
    \hrule
      \includegraphics[width=.45\textwidth]{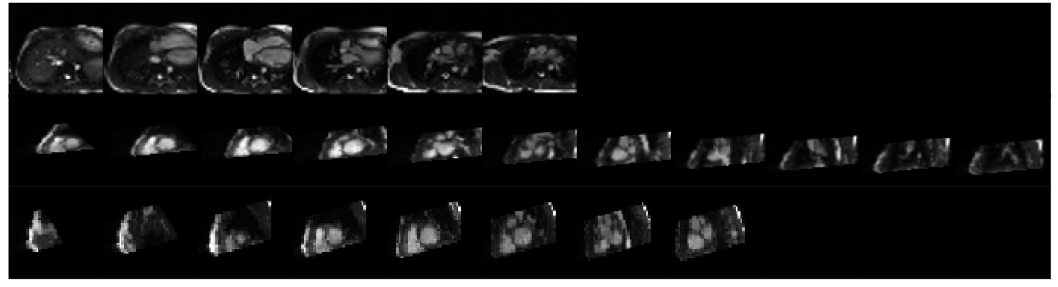}
    \caption{\label{fig:ax2sax_transformations} Three example AX volumes before (upper row) and after the predicted transformation (middle row) in comparison to the AX2SAX ground truth (lower row).
    First example: Good transformation with centered ventricles. The $z$-slices are completely filled with grey level information. Second example: Good transformation leading to a very typical SAX view. Third example (difficult): Thin AX stack. 
    The predicted transformation crops the volume at the apex; even the ground truth transformation is not optimal due suboptimal field of view in the AX acquisition (RV not completely depicted).}
\end{figure}

\begin{figure}
  \centering
      \includegraphics[width=.45\textwidth]{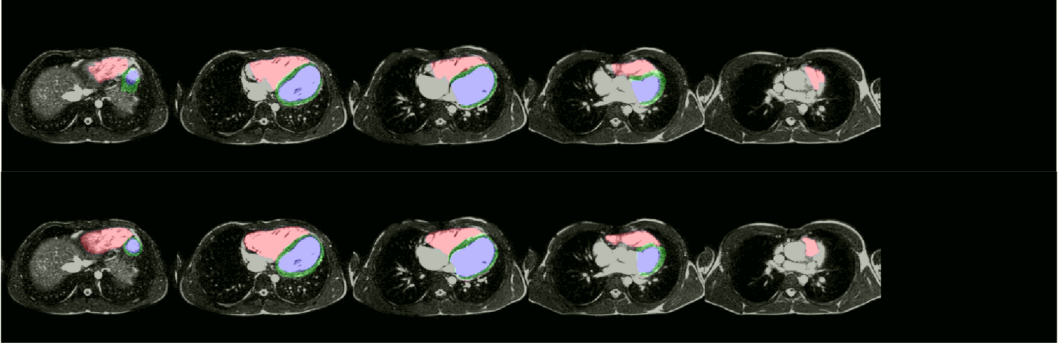}
    \hrule
      \includegraphics[width=.45\textwidth]{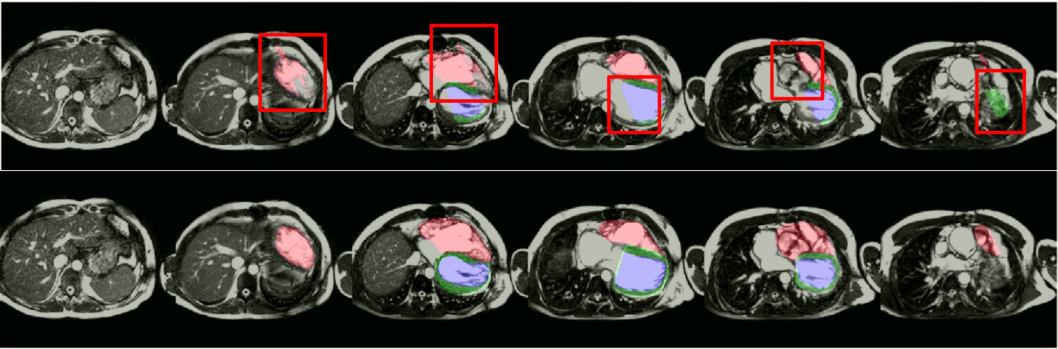}
    \hrule
      \includegraphics[width=.45\textwidth]{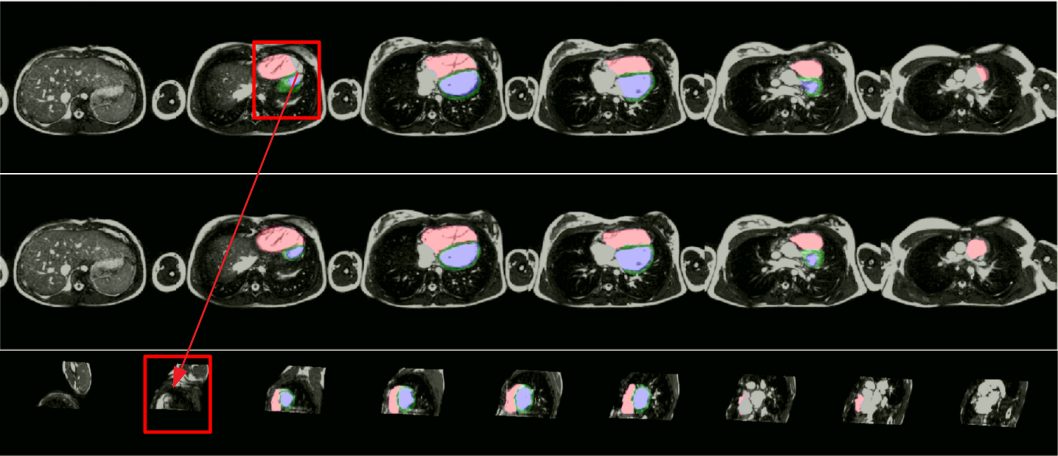}
    \caption{\label{fig:ax_predictions} Three examples of the final predictions overlayed over the axial input image (upper row) vs. the ground truth (lower row). Each row shows six slices, linearly spread over the $z$-direction.
    The first example shows a typical prediction with an average Dice, the second example highlights some common errors of the pipeline. The third example shows a prediction were our segmentation module failed to segment the slices close to the apex, also indicated in the additional third row of this example. Inability to segment the apex is a known weakness of current approaches \cite{Bernard2018a}.}
\end{figure}

Fig. \ref{fig:ax2sax_transformations} shows two good and one sub-optimal example CMR transformation after applying our full pipeline. 
Fig. \ref{fig:ax_predictions} illustrates, beyond that, three example segmentation results when the complete model is applied to an held-out AX image stack, as illustrated in the application scenario (Fig. \ref{fig:application}).
Fig. 4 in the Supplemental Material shows five further cases.
The overall mean Dice / $d_{H}$ of our complete proposed pipeline trained with $Loss$ (Eq.\ref{sec:finalLoss}), that includes the domain adaptation approach, is 
$0.86\pm{0.06}$ / $9.62\pm{3.97}$ for LV, 
$0.65\pm{0.08}$ / $13.72\pm{3.70}$ for MYO and 
$0.77\pm{0.10}$ / $18.31\pm{5.77}$ for RV. 

\begin{figure*}[!p]
\centering
  \includegraphics[width=0.8\linewidth]{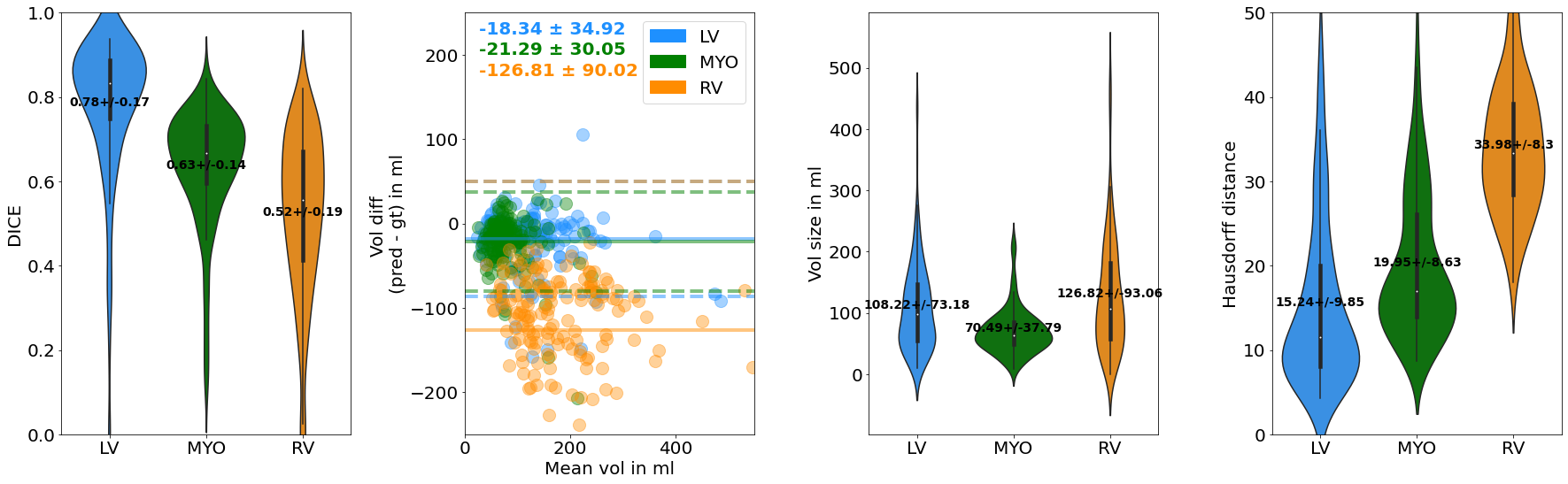}
  \caption{\label{fig:domain_gap}Domain gap when SAX model is applied on axial slices without domain adaptation. From left to right: 1) Violin plot for the mean$\pm{SD}$ Dice. 2) Bland-Altman plot. The mean volume difference between predicted volume and axial ground truth is represented as a solid horizontal line, and the label-wise limits$\pm{1.96 SD}$ are
  represented as dashed lines. 3) Violin plot for the mean$\pm{SD}$ absolute volumes. 4) Violin plot for the mean$\pm{SD}$ Hausdorff distance $d_H$ in mm. Corresponding results in Tab.\ref{table:dice_scores_unet}, third row.}
  
  \includegraphics[width=0.8\linewidth]{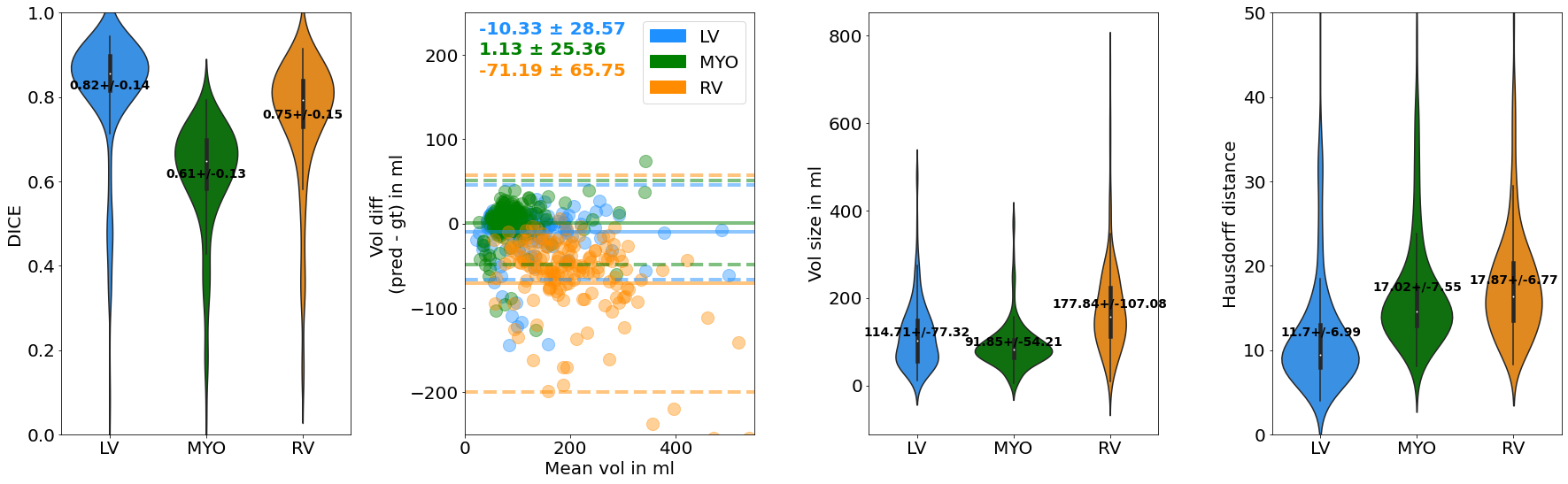}
  \caption{\label{fig:unet_on_resample_ax2sax}Upper limit with respect to a direct regression on the transformation parameters: Ground truth transformation is applied to the AX volumes and the performance of the segmentation module is assessed. Corresponding results in Tab.\ref{table:dice_scores_unet}, last row.}
  
  \includegraphics[width=0.8\linewidth]{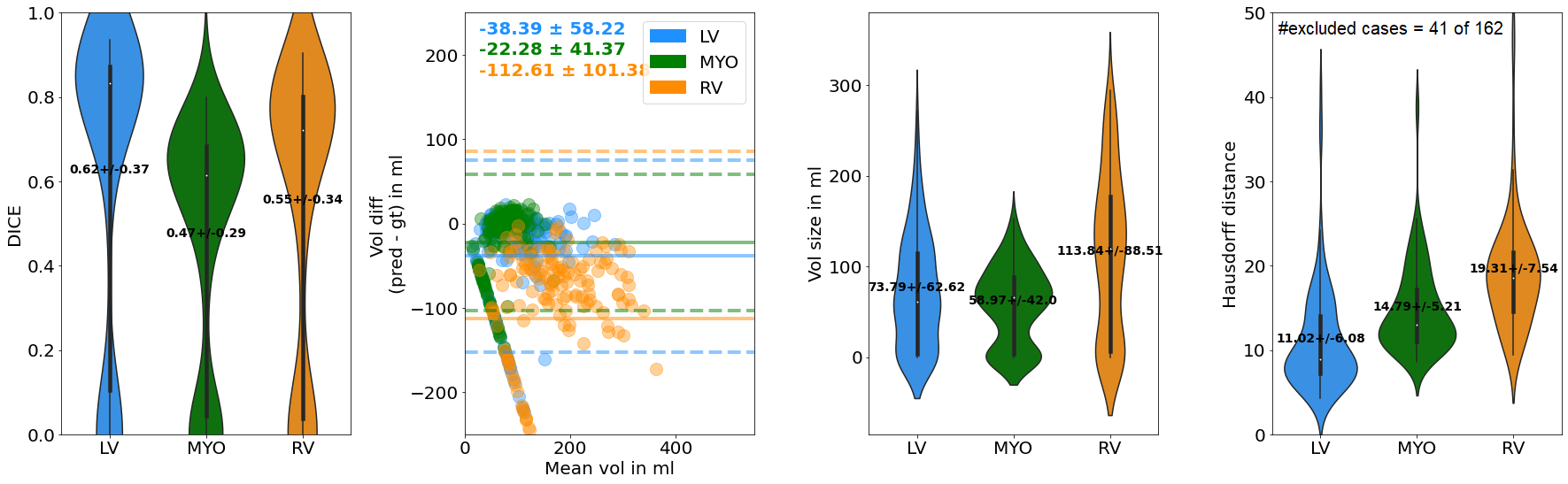}
  \caption{\label{fig:ax1sax0focus0Noinplane}Results of the baseline domain adaptation approach presented in Section \ref{sec:trans_base}), which created reasonable results, but unfortunately the learning process was very unstable and differed from fold to fold. For 41 out of 162 3D volume pairs, the model did not converged at all. These cases were omitted in the computation of $d_H$.  Corresponding results in Tab.\ref{table:dice_scores_per_experiment}, first row.}
  
  \includegraphics[width=0.8\linewidth]{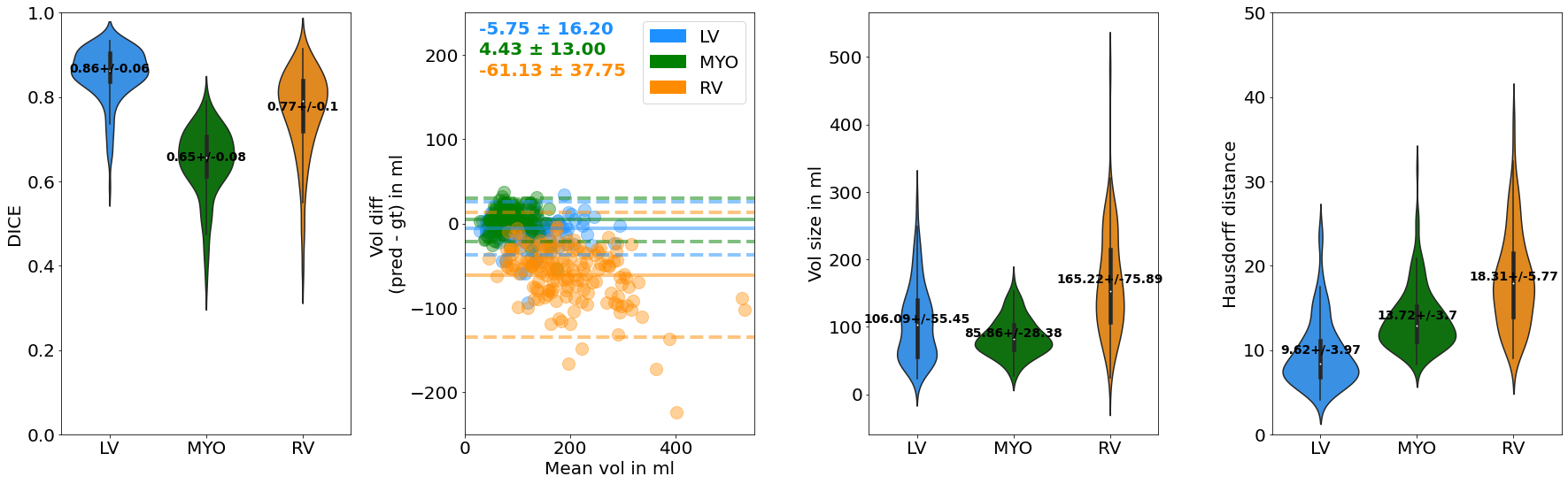}
  \caption{\label{fig:evaluation_TM_pipeline_full}Results of the complete domain adaptation approach using segmentation and transformation module with all extensions. This pipeline considerably improved the Dice and $d_{H}$ score compared to the model without DA, the baseline model and is also better than a direct regression on the transformation parameters. Corresponding results in Tab.\ref{table:dice_scores_per_experiment}, last row.}
\end{figure*}

Fig. \ref{fig:evaluation_TM_pipeline_full} illustrates the corresponding violin and Bland Altman plots. The results show that the domain gap could be considerably reduced and the approach delivers more robust outputs then all other presented versions.

More detailed investigations regarding the areas in which our pipeline functions particularly well or suboptimal reveal that the peripheral areas for the LV and MYO at the apex and base were slightly over-segmented. Fig. 3 of the Supplemental Material shows exemplary the Dice per label and slice of a randomly selected 3D volume. This figure aligns with the visual results in the Bland Altmann plot in Fig. \ref{fig:evaluation_TM_pipeline_full}, where we noticed a mean over-segmentation of $+5.75$ml for the LV and a mean under-segmentation for the RV of $-61.13$ml.

\section{Discussion and Conclusion}
\label{sec:discussion}
We created a pipeline which is capable of segmenting axial CMR from a very heterogeneous multi-site CMR dataset of a rare pathology without the need of any axial contours from this view.
Our pre-trained segmentation module has never seen an axial image stack nor axially contoured data, which shows that labels from other domains can be exploited efficiently (C1). 

In this work, we proposed an unsupervised domain adaptation method, which relies on AX-SAX image pairs to train a novel 3D transformation module in absence of the actual ground truth for the task that should be solved (C1).
As our method neither requires access to SAX nor AX labels and is only trained on paired CMR image stacks, we consider it as a form of unsupervised domain adaptation, according to the common definition of unsupervised domain adaptation provided in Section \ref{sec:introduction}. Training of the external task network is completely independent and can be, in principle, supervised or unsupervised.

Our pipeline is able to adapt data from a domain with less representatives (and in this scenario, no ground truth) into a domain where huge research efforts take place with regard to methodological development, and where data is, relatively spoken, abundant. The rationale behind our domain adaptation approach is to be flexible enough to incorporate well-performing existing and future models trained on SAX views, taking into account the constantly growing openly available CMR data bases.

For now, we injected a pre-trained U-Net for segmentation. 
Other segmentation modules, trained on bigger datasets, different labels or other pathologies could be incorporated without any change of the current pipeline.In general, other task networks like a landmark localizer could theoretically also be incorporated into the current architecture, since $Loss_{Focus}$ is computed on the predicted pixel-wise probabilities of the softmax output (C2). For other tasks like (multi-class) classification, a particular part of the loss function would need to be adjusted in a way that the probability for one class is maximized against the other class(es). However, more experiments are necessary to investigate the performance of the transformation module in complementary tasks.

Owing to the incorporated cycle-consistent loss and focal loss, promising results could be achieved. Comparison of our final results (Table \ref{table:dice_scores_per_experiment}, last row, Fig. \ref{fig:evaluation_TM_pipeline_full}) versus the results obtained from the baseline transformation module (Table \ref{table:dice_scores_per_experiment}, first row, Fig. \ref{fig:ax1sax0focus0Noinplane}) show improvement in Dice and Hausdorff distance for all labels, e.g.  
LV Dice (+24\%),
MYO Dice (+18\%),
RV Dice (+22\%). This indicates that the proposed extensions bring a considerable performance gain (C3). Comparison of our final results versus the application of a supervised SAX U-net on AX (Table \ref{table:dice_scores_unet}, third row, Fig. \ref{fig:domain_gap}) shows that the domain gap could be considerably reduced, especially with regard to the segmentation of the RV. 

The Dice improved by
LV (+8\%),
MYO (+2\%),
RV  (+25\%), while the Hausdorff distance $d_{H}$ could be decreased by
LV (-5,62 mm),
MYO (-6,23 mm),
RV (-15,67 mm) (C3).
Furthermore, the Dice results are also  $2-4\%$ better than the upper limit of a direct regression task (Fig. \ref{fig:unet_on_resample_ax2sax}).
Comparing the segmentation results after domain adaptation  with results that can be obtained by a supervised U-Net on native SAX images, i.e. in a scenario where considerably less domain adaptation is required, (cf. \cite{Bernard2018a} and our results in Tab. \ref{table:dice_scores_unet}, 1st and 2nd row), we can observe a performance gap. So far, our domain adaptation approach has the freedom to account for different slice orientations. In practise, varying voxel spacings, scanner types etc. can also contribute to the domain gap, which was not extensively investigated in our work so far. Note that in previous work \cite{Koehler2020}, we assessed the generalization gap between pediatric congenital cases and cases from adult cardiology on the same overall data sets.

Our domain adaptation pipeline attempts to alter the samples from the target domain to bring the distribution of the target closer to that of the source and (potentially) transform the result back into the target domain. In Toldo \textit{et al.'s} classification \cite{toldo2020unsupervised}, it would belong to the group of \textit{input adaptation}. To achieve that, rigid transformations are estimated. 
In contrast, many current unsupervised DA approaches (e.g., presented in \cite{wilson2018survey,toldo2020unsupervised}) use generative adversarial machine learning techniques to construct a common representation space for the two domains, where feature representations from samples in different domains are encouraged to be indistinguishable.
As mentioned earlier, Self-Training \cite{zou2019confidence} is a research direction of unsupervised DA, which is tightly coupled to semi-supervised learning. Our approach is somewhat related, as we also employ pseudolabels produced by the supervised U-Net to guide the transformation between axial and short-axis ($Loss_{focus}$).

Our approach learnt to extract the patient-specific orientation of the heart on standardized axial slices, and by this, a set of transformation parameters which are necessary to transform the slice orientation to a short-axis view. The \emph{Euler2Affine} layer makes sure that only affine matrices are estimated, which allow valid inverse transformations such that cycle constraints can be imposed.

\begin{figure}
\centering
  \includegraphics[width=\linewidth]{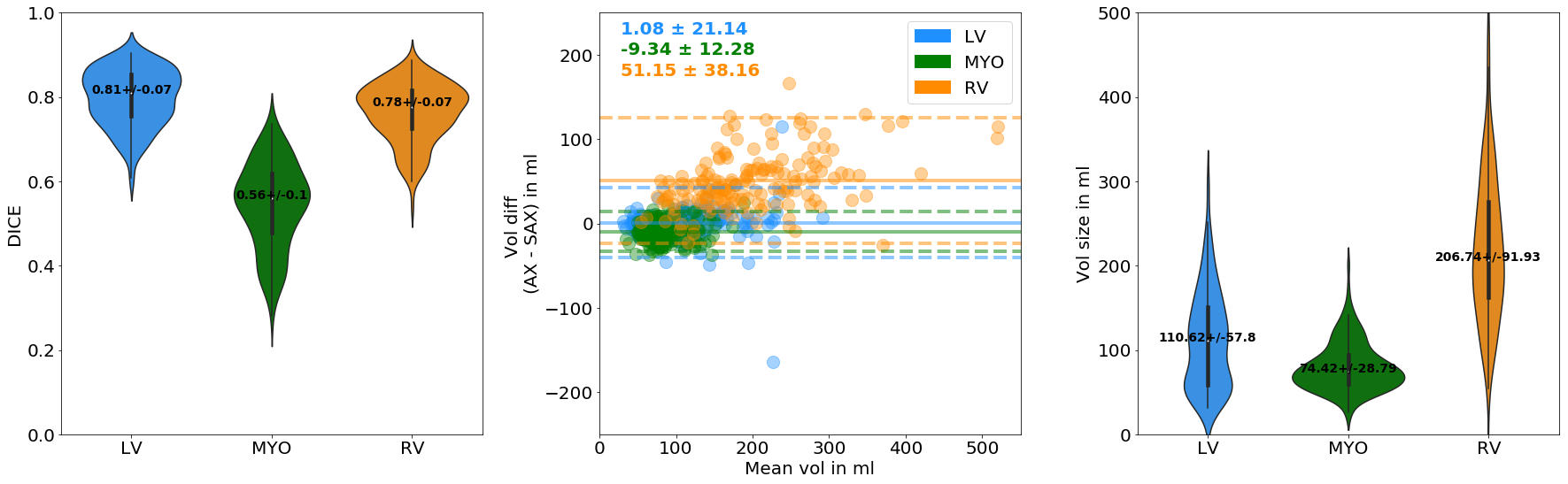}
  \caption{Left: Violin plot for the comparison of median$\pm{SD}$ Dice between the AX and SAX ground truth segmentation. Middle: Bland Altman plot for the volume difference. Right: Violin plot for the median$\pm{SD}$ AX absolute volume.
 }
  \label{fig:ax_sax_alignment}
\end{figure}

Looking at the ground truth, we observed major differences in expert contours of the AX and SAX slice stacks, even though contouring was conducted by the same experienced pediatric cardiologist adhering to the same rules. A similar outcome was shown by other authors of a clinical study \cite{Alfakih2003}.
Fig. \ref{fig:ax_sax_alignment} provides a comparison of the ground truth contours. Interestingly, the average AX RV volume appears $51\pm{15}$ ml larger than in the corresponding average SAX volume. The T-test between RV ground truth volumes (AX-SAX) reveals a score of 5.46227 with $p < 0.01$,
meaning that the difference is significant. Consequently, the delineated shape and volume of RV might vary dependent on the orientation of the $z$-axis to an extent that surpass the expectable differences related to different grid orientations. 

In terms of interpretation of our results, we need to consider this additional point. Our U-Net model was trained on samples from the SAX domain using masks from this domain. However, during testing, the results were compared to masks from different domains, as listed in Tab. \ref{table:dice_scores_unet} and Tab. \ref{table:dice_scores_per_experiment}.  
Therefore, models that have only seen native SAX ground truth contours during the training process might systematically predict smaller RV regions than the expert contour in axial view indicates.

Further limitations of the current approach constitute that the volumes are re-sampled in 3D, potentially increasing uncertainty due to interpolation. Furthermore, in rare cases, the transformation module might cut relevant regions of the structures of interest. 

An interesting route for further investigations is to look in the opposite direction, i.e. for transformation from SAX to AX view. In terms of domain adaptation, this would mean to transform source data (images, labels) to target domain, e.g. to increase the amount of training data in AX orientation. This perhaps enables training of supervised models with axial samples from scratch. The same could be achieved by simply incorporating an own encoder for SAX to AX view transformation into the current architecture. However, we found the current approach most attractive, as available SAX models could be easier integrated. 

To sum up, we proposed an unsupervised domain adaptation approach, which is able to align anisotropic image slice stacks with significantly different fields of view and small overlapping image regions. From the perspective of a world coordinate system, you could think of it as an automatic resampling step with regard to the orientation, where the $z$-direction should be identified. From the perspective of the image space, which is the way we followed in our implementation, it can be rather seen as filling a target grid with slices that match the distribution that an injected 2D task network has previously learnt.
Two major extension to a simple transformation module were proposed: We imposed cycle consistency on forward and backward transformations, which increased the stability of the estimation (C3). This is enabled by constraining the predicted transformations to be invertable.
Furthermore, a loosely coupled focus loss guides the network in optimizing the transformation such that relevant regions are included in the target grid (target with respect to the transformation). 
The method does not require target labels to be optimized (C1) and is also vastly independent to the actual task that should be performed (C2). 
In particular, we showed how image analysis of a data base with rare pathologies can benefit from such an approach that efficiently re-uses existing labels.

\bibliography{references.bib} 
\bibliographystyle{IEEEtran}

\end{document}